\documentstyle[preprint,aps]{revtex}

\def\bra#1{\left\langle #1\right|}
\def\ket#1{\left| #1\right\rangle}

\begin{document}
\draft
%
%\preprint{LA-UR-93-XXX}
\preprint{hep-ph/9405366}
\title{$\pi N$ Scattering in a Large-$N_c$ $\sigma$-Model}

\author{Michael P.\ Mattis\cite{email} and Richard R.\ Silbar}
\address{Theoretical Division, Los Alamos National Laboratory\\
         Los Alamos, NM 87545 USA}

%\date{\today}
\date{May 1994}
\maketitle

\begin{abstract}
\divide\baselineskip by 4
\multiply\baselineskip by 3

We review the large-$N_c$ approach to
meson-baryon scattering, including  recent interesting developments.
We then study $\pi N$ scattering in a
particular variant of the linear $\sigma$-model, in which
the couplings of the $\sigma$ and $\pi$ mesons to the nucleon are
echoed by couplings to the entire tower of $I=J$ baryons (including
the $\Delta$) as dictated by large-$N_c$ group theory. We
sum the complete set of  multi-loop meson-exchange
$\pi N\rightarrow\pi N$  and $\pi N\rightarrow\sigma N$
Feynman diagrams, to leading order in $1/N_c$. The key idea,
reviewed in detail, is that large-$N_c$ allows the approximation
of \it loop \rm graphs by \it tree \rm
graphs, so long as the loops contain at least one baryon leg; trees,
in turn, can be summed  by solving classical equations of motion.
We exhibit the resulting partial-wave $S$-matrix and the rich nucleon
and $\Delta$ resonance spectrum of this simple model, comparing not
only to experiment but also to $\pi N$ scattering in the Skyrme model.
The moral is that much of the detailed structure of the meson-baryon
$S$-matrix which hitherto has been uncovered only with skyrmion
methods, can also be described by models with explicit baryon fields,
thanks to the $1/N_c$ expansion.
\end{abstract}


% body of paper here
\divide\baselineskip by 4
\multiply\baselineskip by 3

\section{Overview of the Large-$N_c$ Approach to Meson-Baryon Scattering}
   \label{sec:introdn}

It is well known\cite{tHooft,Veneziano,WittenI,ColWit,Georgi,Luty}
 that QCD simplifies greatly in the limit
$N_c\rightarrow\infty$, $N_c$ being the number of colors.
Not surprisingly, the large-$N_c$ limit has likewise proved to be very useful
in studying effective low-energy hadron Lagrangians for the Strong
Interactions.
Broadly speaking, such effective theories fall into two categories.
On the one hand, there is the straightforward Feynman diagrammatic approach
in which mesons and baryons are each treated as explicit  dynamical
fields, while on the other hand, there is the more economical
skyrmion picture\onlinecite{Skyrme,ANW}
in which baryons are viewed as solitons constructed
from the meson degrees of freedom.  Since both these approaches
purport to describe the low-energy Strong Interactions, it follows
that if they are sensible, they should be equivalent to one another.
Furthermore, this equivalence must hold order by order in $1/N_c.$
The first steps towards establishing such an equivalence are just
recently being taken\cite{AM,DHM,DiakPet,Japs}.

In either approach, a particularly fruitful physical
process to examine has been
meson-baryon scattering in the large-$N_c$ limit. The present paper
furthers this study, taking as a tractable example of a multi-channel
Lagrangian a  variant of the linear $\sigma$-model. Before
we specify the model, and our particular treatment of it,
it is helpful to put the present work in
historical context.

A review of the
relevant theoretical literature over the past decade reveals an
interesting sociological phenomenon: there are two disjoint bodies of
large-$N_c$ papers
devoted to two topologically distinct sets of diagrams,
namely \it Compton-type \rm versus \it exchange-type \rm graphs, that
contribute to the meson-baryon $S$-matrix.\footnote{
\divide\baselineskip by 2
So  far as we are aware, the only attempt to date to treat these
two classes of graphs in a unified manner can be found in Sec.~7 of
Ref.~\onlinecite{DHM}.}
 Examples of Compton-type and
exchange-type graphs are displayed in Fig.~1 and Fig.~2, respectively.
Topologically, they differ in the following way: in the exchange-type
graphs of Fig.~2, it is possible to trace a continuous line from the incoming
to the outgoing meson without ever traversing a baryon line segment,
whereas in the Compton-type graphs of Fig.~1 this cannot be done.

Let us review,  briefly, some of the salient points of physics that emerge
from the study of each of these two classes of diagrams.

   \subsection{ Compton-type graphs }

While presently the Compton-type graphs (Refs.~
\onlinecite{DiakPet,Japs,GerSak,DashMan,DashManII,DashJenMan,Jenkins})
 are much less well understood than the
exchange-type graphs discussed below, they nevertheless yield some
interesting physics, as follows.
Look at Figs.~1a and 1b. Since each vertex
scales like $\sqrt{N_c}$ (see Ref.~\onlinecite{WittenI}),
 these graphs  individually scale like\footnote{
\divide\baselineskip by 2
The  baryon propagator is approximated by
$i(v\cdot k+i\epsilon)^{-1}$ in the large-$N_c$ limit, where $v$
is the baryon's 4-velocity, $k$ is the momentum imparted by
the incoming meson (assuming the incoming baryon to be on shell),
and it is also understood that one throws away the two small
components of the Dirac 4-spinor.
We focus on the kinematic regime  $k\sim N_c^0$ so that the baryon propagators
 do not affect the $N_c$ counting.}
$N_c.$
However, we know from Witten's analysis of quark-gluon
diagrams\onlinecite{WittenI} that the total amplitude for
meson-baryon scattering must scale like $N_c^0,$ not
$N_c.$ Therefore there must be leading-order cancellations between
Figs.~1a and 1b. Add to this observation another
important piece of large-$N_c$ physics: the fact that for the case
of two light flavors (which we focus on exclusively herein) the
spectrum of stable baryons is a tower of states of equal spin and
isospin\onlinecite{ANW}:
$I=J=1/2,3/2,5/2,\cdots,N_c/2$, which are
 all degenerate in the large-$N_c$ limit
(more precisely, in the limit $J^2/N_c\rightarrow0$). We then demand
leading-order cancellation between Figs.~1a and 1b, for the reason
described above, with the three baryon legs drawn from  all possible
baryon states in the $I=J$ tower, consistent with triangle inequalities
for isospin and angular momentum at each vertex. This exercise is
carried out in Refs.~\onlinecite{GerSak} and \onlinecite{DashMan}.
The upshot is a set of proportionality relations between the various
coupling constants $g_{\pi NN},$ $g_{\pi N\Delta},$ $g_{\pi \Delta\Delta},$
and so forth up the $I=J$ tower, relating each of these \it a priori \rm
independent couplings to a single underlying coupling constant,
up to multiplication by Clebsch-Gordan coefficients.
We call this set of relations for the pion-baryon couplings
the ``proportionality rule.'' Furthermore, Dashen and Manohar have
shown that corrections to the proportionality rule do not occur
at order $1/N_c,$ as naively expected, but rather at order
$1/N_c^2$\onlinecite{DashManII}. This suggests that the proportionality rule
should be relatively robust. Calculationally, it implies that,
once the order $N_c$ contributions to the amplitude have cancelled,
the surviving order $N_c^0$ pieces arise solely from the
$1/N_c$ corrections to the baryon propagator, and not from
$1/N_c$ corrections at the vertices, as one might have thought.

Numerically, the proportionality rule for the pion-baryon couplings
works well. Not only does the decay width of the $\Delta$ work out to
within a few GeV of its measured value when $g_{\pi N\Delta}$ is
related, using this rule,
 to the experimental value of $g_{\pi NN}$\onlinecite{ANW,DHM};
but furthermore, with the same input parameters, the widths of the
``large-$N_c$ artifacts,'' \it i.e. \rm the baryons with $I=J\ge5/2,$
are so large that they cannot be considered ``particles'' at all, and
as such, pose no problem for
phenomenology\onlinecite{DHM}. This latter observation
removes what has been, till recently, one of the chief objections to the entire
large-$N_c$ program. Another success of large $N_c$ is that the
group-theoretic predictions of the old $SU(2N_F)$ symmetry are
recaptured\onlinecite{Georgi,Luty,DashMan,DashJenMan}, without
the need to appeal to the construct of the nonrelativistic,
weakly interacting constituent quark model.

A further refinement was made recently by Jenkins\onlinecite{Jenkins},
who examined the one-loop chiral corrections to the masses
$M_J$ of the $I=J$ baryons, and deduced the consistency relations
\begin{equation}
 M_J\ =\ M_0\ +\ {J(J+1)\over2{\cal I}}\ +\ {\cal O}(N_c^{-2})
\label{Jenkinseqn}
\end{equation}
where $M_0$ and $\cal I$ are constants of order $N_c$ that can
be fixed, for example, by pegging $M_{1/2}$ and $M_{3/2}$, respectively,
to the experimental nucleon and $\Delta$ masses.

While the large-$N_c$ results of
Refs.~\onlinecite{DashMan,DashManII,DashJenMan,Jenkins} are derived
using effective Lagrangians of mesons and explicit baryons,
the physics of the Compton-type graphs can also be accessed using
the skyrmion approach\onlinecite{DiakPet,Japs}. The parallelism between
the two approaches is manifest in expressions such as Eq.~(\ref{Jenkinseqn}).
In the language of the two-flavor Skyrme model, $M_0$ and $\cal I$
are interpreted as the mass and moment of inertia of the soliton,
respectively\onlinecite{ANW,Schulman}. It is reassuring that
the expression (\ref{Jenkinseqn}) can also be gotten directly
from looking at quark diagrams in
large-$N_c$ QCD\onlinecite{Georgi,Luty}, closing the circle.

\subsection{Exchange-type graphs}

Next we turn to the physics of the exchange-type graphs
(Refs.~\onlinecite{AM,HEHW,Sig,MandK,MandP,Karliner,ninj,Donohue,Muk,Action}),
which is the primary focus of this paper. Examples are shown in Fig.~2.
These graphs likewise contribute to the
scattering amplitude starting at order $N_c^0.$ Although the
summation of \it all \rm such graphs would appear to be an impossible task,
it can actually be carried out in a straightforward manner---so
long as one contents oneself with the leading-order answer in the
$1/N_c$ expansion\onlinecite{AM}.  As will be reviewed in detail
in the Sections to follow, the key idea is to rewrite these
multiloop graphs as \it trees\rm, exploiting the large-$N_c$
approximation. Tree graphs have the great
advantage over loops that they can all be summed by solving \it classical \rm
equations of motion.\footnote{
\divide\baselineskip by 2
To remind the reader\onlinecite{AM}
 that he or she already knows
a situation where ``loops'' become ``trees,'' recall the ancient problem
of electron-proton scattering in the low-energy regime where the proton
mass is much greater than all other scales in the problem. On the one
hand, these are evidently
multiloop interactions, in which the proton and electron
lines exchange a large number of photons in all possible tanglings. On
the other hand, we know that
the physics is accurately described by \it classical \rm equations:
first the proton generates a classical Coulomb field, and then the
electron propagates linearly through this non-trivial
 background (Rutherford scattering). These
two disparate pictures are reconciled by the fact that
the loop graphs are really trees, by
exactly the same manipulations described in Sec.~II below. The insight of
Ref.~\onlinecite{AM} is that this same mechanism (modulo nonlinearities
due to the fact that bosons, unlike photons, are self-interacting)
holds for the exchange of arbitrary bosons in the large-$N_c$ limit, thanks to
the  proportionality rule as well as the $I_t=J_t$ rule reviewed below.}
It is this summability property which justifies
our earlier statement that the exchange-type graphs are much better
understood than the Compton-type graphs.

 While the analysis of this paper will be carried
out using explicit baryon fields, the set of classical
equations that emerges is, once again, highly reminiscent of the
skyrmion approach, in which the corresponding
classical equations describe a pion propagating through
 the background field generated by the skyrmion
itself\onlinecite{HEHW,Sig,MandK,MandP,Karliner,ninj}.
In particular, the group-theoretic relations familiar from the Skyrme
model carry over intact to models such as the present one
with explicit  baryons.
These include non-trivial, and experimentally reasonably well satisfied,
relations in which isospin-$3/2$ $\pi N$ scattering amplitudes are
expressed as linear combinations of the  isospin-$1/2$
amplitudes\onlinecite{HEHW,MandP}. Similar relations hold for
kaon-nucleon scattering \onlinecite{Karliner}, and for $\pi N\rightarrow
\rho N$\onlinecite{ninj}, and in fact for all
quasielastic meson-baryon scattering processes.

If, extending
Donohue's original suggestion\onlinecite{Donohue}, one crosses these
 relations among scattering
amplitudes from the $s$-channel
to the $t$-channel (\it e.g.\rm, $N\bar N\rightarrow\,$mesons),
they can be re-expressed concisely
as two large-$N_c$ selection rules\onlinecite{Muk,Action}.
 First, there is the very same
``proportionality rule'' discussed earlier, in the context of the
Compton-type graphs. However, the derivation given in
Ref.~\onlinecite{Muk} makes clear that
the proportionality rule is \it completely independent
of the chiral limit\rm, and furthermore, that it applies not only to the
pion-baryon couplings but equally to the baryon couplings of
\it all \rm bosons. Beyond the width calculations noted
above\onlinecite{ANW,DHM}, the
phenomenological validity of the proportionality rule is put to the
test in Fig.~7 of Ref.~\onlinecite{MandP}, in which the appropriate
linear combinations of the experimental
$\pi N\rightarrow\pi N$ and $\pi N\rightarrow\pi\Delta$ scattering
amplitudes are compared.

 In addition, a second large-$N_c$ selection rule emerges, the ``$I_t=J_t$
rule''\onlinecite{Muk,Action}. This rule states that
the isospin of the emitted/absorbed meson must equal its \it total \rm
(spin + orbital) angular momentum, measured
in the rest frame of the large-$N_c$
baryon. Concrete examples of meson-nucleon couplings that
satisfy the $I_t=J_t$ rule include the pseudovector coupling of the
pion, the tensor coupling of the $\rho,$ and the vector coupling of
the $\omega$ \onlinecite{Action}:
\begin{equation}
 \big(g_{\pi NN}/2M_N\big)\partial_\mu\vec\pi\cdot\bar N
\gamma^5\gamma^\mu\vec\tau N\ ,
\quad
 g^{\rm tens}_\rho\partial_\mu\vec\rho_\nu\cdot\bar N
\sigma^{\mu\nu}\vec\tau N\ ,
\quad
 g^{\rm vec}_\omega\omega_\mu\cdot\bar N\gamma^\mu N\ ,
\end{equation}
each of which must be
augmented by couplings to the entire tower of $I=J$ baryons as
required by the proportionality rule.
Since these couplings obey the $I_t=J_t$ rule, the three coupling constants
are nonvanishing at leading order in the large-$N_c$ expansion:
\begin{equation}
{g_{\pi NN}\over2M_N}\ \sim\  g^{\rm tens}_\rho \sim\
 g^{\rm vec}_\omega\ \sim\ \sqrt{N_c}\ .
\label{nonvanishing}
\end{equation}
In contrast, the $I_t=J_t$ rule \it forbids \rm at leading order
 the other two canonical vector-meson
 interactions, the \it vector \rm coupling of the $\rho$ and
the \it tensor \rm coupling of the $\omega,$
\begin{equation}
 g^{\rm vec}_\rho\vec\rho_\mu\cdot\bar N
\gamma^{\mu}\vec\tau N\quad \hbox{and}\quad
 g^{\rm tens}_\omega\partial_\mu\omega_\nu\cdot\bar N\sigma^{\mu\nu} N\ ,
\end{equation}
meaning that these coupling constants must be down by (at least)
one power of $1/N_c$ compared to Eq.~(\ref{nonvanishing}):
\begin{equation}
 g^{\rm vec}_\rho \sim\  g^{\rm tens}_\omega\ \sim\ {1\over\sqrt{N_c}}\ .
\end{equation}
The relative unimportance of the vector (tensor) coupling of the
$\rho$ ($\omega$) has long been  known to  nuclear physicists who construct
one-boson exchange models of the nucleon-nucleon
potential\onlinecite{LeeTab,BarEb,Bonn,Paris}. It
is pleasing to see these phenomenological
rules of thumb emerge as  theorems in the large-$N_c$ limit.

\subsection{Two interesting unresolved questions}

In lieu of a Conclusions section, we close this expanded Introduction
 with two questions that are food for further thought.
First, is the complete meson-baryon
$S$-matrix properly obtained by adding the Compton-type and exchange-type
graphs together, or, as an alternative prescription, might it be the case
 that either set of graphs \it by itself \rm
(assuming an infinite spectrum of mesons) contains the complete answer?
This latter possibility is suggested by the observation that mesons and
baryons are composite particles made up from quarks and gluons. Since
at the quark-gluon level there is no longer a topological distinction between
the graphs of Fig.~1 and Fig.~2,  one must  be especially careful to
avoid double counting, and this might conceivably preclude adding the
graphs of Fig.~1 and Fig.~2 together in a naive way.\footnote{
\divide\baselineskip by 2
For the resolution of similar issues in atomic physics, namely
 the avoidance of double-counting when bound states are involved, see
Ref.~\onlinecite{Lynn} and references therein.}

Second, the exchange-graph formalism of Ref.~\onlinecite{AM} applies
not only to the meson-baryon system which we focus on here,
but equally to the baryon-baryon, baryon-antibaryon,
baryon-baryon-baryon, and in general to all $n$-baryon, $m$-antibaryon
interactions (Fig.~3). Of course, there are no analogs of
Compton-type graphs for these multi-baryon systems. It follows that
the exchange-graph formalism of Ref.~\onlinecite{AM}
gives---in principle---{\it the complete answer}
 for these cases, to leading order in $1/N_c.$ By this we specifically
mean the following: given an effective hadron Lagrangian
whose meson-baryon couplings properly
embody the $I_t=J_t$ and proportionality rules,
the complete set of Feynman diagrams of the sort exhibited in
Fig.~3 can be summed to leading order in $1/N_c.$
 It would be an interesting exercise
to carry out this program, starting from a well-motivated effective Lagrangian,
 and to compare the results to the popular Bonn\onlinecite{Bonn} and
 Paris\onlinecite{Paris} potential
models (which are derived from just the ladder diagrams with
at most one crossing)  as well as to the recent work of
Weinberg and others that relies exclusively on chiral perturbation
theory\onlinecite{Weinbergchi}.

 \subsection{Outline of paper}

The remainder of this paper is organized as follows. In Sections II
and III we review the exchange-graph
formalism of Ref.~\onlinecite{AM} and apply it to two warm-up  problems,
a ``$\sigma$-only'' and a ``$\pi$-only'' model, respectively.
Sections IV-VI explore in detail the meson-baryon $S$-matrix in
a richer  model comprising both pions and  $\sigma$ mesons, a variation on the
 Gell-Mann-Levy $\sigma$-model\onlinecite{linsigmod}.
 Obviously, we do not take this model seriously as a
realistic depiction of hadron physics. Rather,
we aim only to illustrate how the formalism of Ref.~\onlinecite{AM}
leads in a concrete way to a quantitative calculation of the
 exchange-graph contribution to the  multi-channel meson-baryon
$S$-matrix. With the present model solved,
the scene is  set for more ambitious, realistic calculations,
necessarily incorporating vector mesons.
We are also interested in  comparing the large-$N_c$
effective Lagrangian approach that uses explicit
baryons,  with earlier large-$N_c$ results from the Skyrme model.
We come to the conclusion that much of the detailed structure
of the meson-baryon $S$-matrix which hitherto has been
uncovered only with skyrmion methods, can equally be described by
models with explicit baryon fields. At the same time, both
approaches share significant problems in the low partial waves,
the complete resolution of which remains a major technical hurdle.

\section{First warm-up problem: a $\sigma$-only Model}
   \label{sigonlymodel}

As a first calculation, let us consider a model with only $\sigma$ mesons and
(non-strange) baryons \cite{AM,Boulder}.
Because the $\sigma$ has $I=J=0$, this toy model avoids
the spin and isospin complications due to
non-commutativity of Pauli matrices at the meson-nucleon vertices.
It also avoids the complications of inelastic 2-body channels ($e.g.,$
nucleons cannot turn into $\Delta$'s).

The Lagrangian  to be solved in this Section is the large-$N_c$ version of:
\begin{equation}
   {\cal L}_{\sigma N} =
   {1 \over 2} (\partial_\mu \sigma)^2 - V(\sigma)  +
   \overline{N} (i \gamma\cdot\partial - M_N) N -
   g \sigma \overline{N} N \  ,
\label{sigLagn}
\end{equation}
where, for definiteness, the $\sigma$ self-interactions are
described by the fourth-order potential
\begin{equation}
   V(\sigma) = {1 \over 2} {m_\sigma}^2 \sigma^2 +
   {1 \over 6} \kappa \sigma^3 +
   {1 \over 24} \lambda \sigma^4 \  .
\label{sigPotl}
\end{equation}
By the words ``large-$N_c$ version of'' we mean that the
coupling of the $\sigma$ to the nucleon in
Eq.~(\ref{sigLagn}) must, in principle, be augmented by analogous
couplings to the entire $I=J$ tower of large-$N_c$ baryons, starting
with the $\Delta$ ($I=J=3/2$) and continuing through the state with
$I=J=N_c/2$. The relative strengths of these couplings is given by
the proportionality rule \cite{Muk}.
However, in this simple model, since the
$\sigma$ carries the quantum numbers of the vacuum, it couples
 diagonally to this tower (as noted above).
 Therefore, so long as we restrict our
attention to  nucleon targets, we can safely drop these additional couplings
to the higher baryon states
and work  with the simplified Lagrangian (\ref{sigLagn}).

In the large-$N_c$ limit the nucleon has mass of order $N_c$ and its
degrees of freedom freeze out. This means that the nucleon
kinetic energy term in Eq.~(\ref{sigLagn}) can be dropped, and the Yukawa
term has $\overline{N} N$ replaced by a static source $j({\bf x})$.
The formal derivation of this  intuitive prescription was given in
Ref.~\cite{AM}. For completeness, we  review it here. Looking at
Fig.~4, the product of the nucleon propagators (reading from bottom to
top) is
\begin{eqnarray}
        {i\over {p\llap/}+{k\llap/}_1-M_N^{}+i\epsilon } & &\times
        {i\over {p\llap/}+{k\llap/}_1+{k\llap/}_2-M_N^{}+i\epsilon }
        \times\cdots\times
        {i\over {p\llap/}+{k\llap/}_1+\cdots+{k\llap/}_{n-1}-M_N^{}+i\epsilon }
        \nonumber \\
        & &\approx\
        {{\gamma_0+1} \over 2} {i \over k_{10}+i\epsilon} \times\cdots\times
        {i \over k_{10}+\cdots+k_{n-1,0}+i\epsilon} \ .
\label{PropProd}
\end{eqnarray}
In the above we have taken the large-$N_c$ (\it i.e.\rm,  nonrelativistic)
limit of the nucleon propagators
%
%\eqn\eqa
\begin{equation}
{i\over {p\llap/}+{k\llap/}-M_N^{}+i\epsilon }\quad
{\stackrel{\scriptstyle{N_c\rightarrow\infty}}{\longrightarrow}}\quad
{\gamma_0+1\over2}{i\over k_0+i\epsilon} \ ,
\label{eqa}
\end{equation}
assuming that the nucleon is in its rest frame.
The prefactor $(\gamma_0+1)/2$ is the projector onto the large components
of the Dirac 4-spinor.  From now on we suppress it, with the understanding
that we always throw away the small components.

Our desired result is obtained by summing over the $n!$ crossed ladders
(Fig.~5), and using the interesting identity for distributions,
%
%\eqn\eqb
\begin{eqnarray}
2\pi\delta \big( \sum_{i=1}^nk_{i0} \big) \!\!\!
\sum_{{\rm permutations}\atop{(i_1,\cdots,i_n)}} \
& &\!\!\!\!\!\!\!\!{i\over k_{i_10}+i\epsilon}  \times
{i\over k_{i_10}+k_{i_20}+i\epsilon}
\times\cdots\times
{i\over k_{i_10}+\cdots+k_{i_{n-1}0}+i\epsilon} \nonumber \\
&=&
2\pi\delta(k_{10}) \times 2\pi\delta(k_{20}) \times\cdots\times
2\pi\delta(k_{n0}) \ .
\label{eqb}
\end{eqnarray}
(To prove this,
 Fourier transform both sides of this identity in all $n$ momenta.)
Each of the $n!$ terms in this sum corresponds to a distinct crossing
or ordering of the $n$ rungs of the ladder.
The $\delta$-function on the left-hand side of this
equation reflects conservation of energy along the nucleon line in
the large-$N_c$ limit:
%
%\eqn\eqc
\begin{equation}
{2\pi\delta\big(-p_0'+p^{}_0+\sum_{i=1}^nk_{i0}\big)\quad
{\stackrel{\scriptstyle{N_c\rightarrow\infty}}{\longrightarrow}}\quad
2\pi\delta\big(\sum_{i=1}^nk_{i0}\big)\ .}
\label{eqc}
\end{equation}

Recognizing $2\pi\delta(k_0)$ as the 4-dimensional Fourier transform of
$\delta^3({\bf x}),$ we immediately understand the meaning of the
simple factorized right-hand side of Eq.~(\ref{eqb}) in terms of graphs.
Simply put, the sum of the $n!$ crossed ladders is equal to the
{\it single} graph of Fig.~6, generated by the effective Lagrangian
%
%\eqn\eqd
\begin{equation}
{{\cal L}_{\rm eff}\ =\ {1\over2}(\partial_\mu\sigma)^2
-V(\sigma)-\sigma j({\bf x})}
\label{eqd}
\end{equation}
where, as promised, the nucleon field has been frozen out in favor
of the external $c$-number source
%
%\eqn\eqe
\begin{equation}
{j({\bf x})\ =\ g\,\delta^3({\bf x})\ .}
\label{eqe}
\end{equation}

The complete exchange-graph contribution to
 $\sigma N$ scattering in the large-$N_c$
limit now emerges from a two-stage numerical program,
which is most transparent in graphical terms.  In the first stage,
one defines a ``classical'' $\sigma$ field $\sigma_{\rm cl}$ as the sum of
all one-point trees (Fig.~7).  The reason one considers only trees is
that meson loops are suppressed by powers of
 $1/N_c$\onlinecite{Veneziano,WittenI,Luty}. In the second stage,
one considers a propagating $\sigma$ field (which we call the ``quantum''
field $\sigma_{\rm qu}$ to distinguish it from $\sigma_{\rm cl}$) interacting
with an arbitrary number of $\sigma_{\rm cl}$ insertions (Fig.~8). By
inspection, this two-stage procedure is equivalent to summing
all the tree graphs of the form shown in Fig.~6 (the loop graphs being
subleading in $1/N_c$).  As promised: the  loops (Figs.~4-5) have turned into
trees, exactly as in the old electron-proton problem invoked in Sec.~I.

This two-stage graphical procedure is easily translated into the language of
differential equations. Solving for $\sigma_{\rm cl}$ as per Fig.~7 is
equivalent to solving the classical Euler-Lagrange equation for the
effective Lagrangian of Eq.~(\ref{eqd}), namely,
%
%\eqn\eqf
\begin{equation}
-\nabla^2 \sigma_{\rm cl}({\bf x}) +
V'\big(\sigma_{\rm cl}({\bf x})\big) +  j({\bf x}) \ =\  0\ .
\label{eqf}
\end{equation}
Note that $\sigma_{\rm cl}$ is time-independent because the source
$j({\bf x})$ has
this property. Next, solving for the propagating field $\sigma_{\rm qu}$,
as given by Fig.~8, is accomplished by noticing that at every vertex, there
are exactly two $\sigma_{\rm qu}$ legs, the rest being insertions
of $\sigma_{\rm cl}$, with the coupling constants read off from $V(\sigma)$.
Therefore, the relevant equation of motion comes from
the \it quadraticized \rm Lagrangian
%
%\eqn\eqg
\begin{equation}
{\cal L}_{\rm quad}\ =\ {1\over2}\partial_\mu\sigma_{\rm qu}
 \partial^\mu \sigma_{\rm qu} -{1\over2}\sigma_{\rm qu}^2V''\big(
\sigma_{\rm cl}({\bf x})\big)\ ,
\label{eqg}
\end{equation}
which induces the linear time-dependent equation
%
%\eqn\eqh
\begin{equation}
\big[\partial_\mu \partial^\mu + V''(\sigma_{\rm cl}({\bf x}))\big]
\ \sigma_{\rm qu}(x)\ =\ 0\ .
\label{eqh}
\end{equation}

In short, we have outlined a two-stage numerical procedure,
the first stage involving
a non-linear time-independent equation for a ``classical'' meson field,
the second involving a linear time-dependent equation for a ``quantum''
meson field in the classical background.
This is reminiscent of the skyrmion approach to meson-baryon
scattering\onlinecite{HEHW,Sig,MandK,MandP}.
In the subsequent Sections, when pions are introduced,
this correspondence will be sharpened by the
emergence of a hedgehog structure to the classical pion field that is
familiar from the Skyrme model\onlinecite{Skyrme,ANW}. (The chief
\it difference \rm between the two approaches is, of course, that
baryon number is carried by topology in the Skyrme model, and by
smeared $\delta$-function sources when the baryon fields are explicit.)

The analog of the hedgehog Ansatz in
the present model with $I=0$ $\sigma$ mesons alone is just ordinary
spherical symmetry:
\begin{equation}
  \sigma_{\rm cl}({\bf x}) \equiv G(r)\ .
\label{sigradial}
\end{equation}
The profile
function $G(r)$ is found by solving the non-linear radial
differential equation
\begin{equation}
   G'' +{2\over r}G'- {m_\sigma}^2 G
       - {\kappa \over 2}\,{G^2}
       - {\lambda \over 6}\,{G^3} =
   j({\bf x})
\label{sigGeqn}
\end{equation}
implied by Eqs.~(\ref{sigPotl}) and (\ref{eqf}).
 Unfortunately,  Eq.~(\ref{sigGeqn}) suffers from
ultraviolet problems when $j$ is literally
taken to be a $\delta$-function as per Eq.~(\ref{eqe}).
The source of these divergences (which are worse than in the original
loop graphs, Figs.~4-5)  can be traced to the nonrelativistic
reduction of the propagator (\ref{eqa}), which is only valid so long
as the components of the exchanged meson momentum satisfy $|k_\mu|\ll M_N.$
(A similar breakdown of the large-$N_c$ approach is discussed in
Sec.~8 of Ref.~\onlinecite{WittenI}.)
A simple cure is to smear out the source, say, as a Gaussian:
\begin{equation}
   j({\bf x})\ \longrightarrow\
      {g \over {(a_N \sqrt\pi)^3}} \exp(-r^2/a_N^2) \ ,\quad
a_N^{}\sim N_c^0\ .
\label{gaussian}
\end{equation}
This approximation now renders Eq.~(\ref{sigGeqn}) tractable, at the expense of
introducing a ``nucleon size'' parameter $a_N$ into the problem.
This new parameter provides
an ultraviolet cutoff on the momentum allowed to flow
into or out of the nucleon.
We have checked that our numerical results are not
overly sensitive to $a_N$ over a reasonable range of values.

Equation (\ref{sigGeqn})
represents a two-boundary value problem which can be solved
in an iterative fashion using
a standard ``shoot and match'' Runge-Kutta integration procedure
\cite{NumRec}.\footnote{
\divide\baselineskip by 2
For details, see the Appendix.}
Fig.~9 shows the profile function $G(r)$ for the specific choice of
parameters $m_\sigma=600\,$MeV, $\kappa=18.5$, $\lambda=214$, $g=13.6$
and $a_N=0.5$ fm.
Note that $G(r)$ looks very much like a Yukawa function,
$\exp(-m_\sigma r)/r$, except that it is finite at the origin
(because of the smearing of the nucleon source term) and
has small deviations in the 0.5 to 1.0 fm region due to the non-linear terms
involving $\kappa$ and $\lambda$.

Given $G(r)$, we then solve Eq.~(\ref{eqh}) for
$\sigma_{\rm qu}$ by means of a standard partial wave analysis.
For angular momentum $l$, the radial scattering wave
function $u_l(r) = r \sigma_l(r)$ having energy $\omega$ satisfies
\begin{equation}
  \left[{{d^2}\over{dr^2}}  + q^2 - \kappa G(r) -
    {\lambda\over2} G^2(r) - {l(l+1) \over {r^2}}\right] \, u_l(r) = 0 \ ,
\quad
q^2 = \omega^2 - m_\sigma^2\ .
  \label{sigscat}
\end{equation}
This is a Schr\"odinger-like linear differential equation that can
also be solved by Runge-Kutta integration from the
origin (where $u_l(r)$ must be regular, going like $r^{l+1}$).
The asymptotic form of $u_l(r)$ is then  fit in the usual way
to a linear combination of
spherical Ricatti-Bessel functions, $j_l(qr)$ and $n_l(qr)$, yielding
the  phase shifts for $\sigma N$ elastic scattering.

The potential in Eq.~(\ref{eqh}) [or Eq.~(\ref{sigscat})]
has a short-range repulsive core
(coming from the quartic term in the Lagrangian) and intermediate-range
attraction (coming from the cubic term and the fact $G(r) < 0$).
%
%The coupling constants $\kappa$ and $\lambda$ can be fixed by an argument
%similar to the appeal to the linear $\sigma$-model used in
%Sec.~\ref{sec:sigpimodel}.
%
Consequently, as shown in Fig.~10, the $S$-wave phase
shift at low energies is positive because of the medium-range
attraction, but it soon turns over and looks like the phase shift for
a hard-core repulsive potential.
At still higher energies (not shown), the phase shift returns to 0, since the
short-range repulsive core is finite. The higher partial waves exhibit
similar behavior, but offset to increasingly higher energies because of the
angular momentum barrier.

\section{second warm-up problem: a $\pi$-only model}
   \label{pionlymodel}

As a second simplified
 example, we consider a model of gradient-coupled pions and
$I=J$ baryons.  The Lagrangian we want to solve is the large-$N_c$ version
of
\begin{eqnarray}
   {\cal L}_{\pi N} &=&
   {1 \over 2} \partial_\mu \pi^a \; \partial^\mu \pi^a
   - {1 \over 2} m_\pi^2 \; \pi^a \pi^a
   - {\lambda \over 24} ( \pi^a \pi^a )^2 \nonumber \\
   & & \quad + \overline{N} (i \gamma\cdot\partial - M_N) N
   -g_\pi\partial_\mu\vec\pi\cdot
   \overline{N}\gamma^5\gamma^\mu\vec\tau N\ .
\label{piLagn}
\end{eqnarray}
The reason for choosing pseudovector coupling rather
than the pseudoscalar coupling,
$-g'_\pi\vec\pi\cdot\overline{N}i\gamma^5\vec\tau N$, is that it
is more amenable to a large-$N_c$ treatment, for the following
reason.
The matrix $\gamma^5$ is purely off-diagonal, connecting the large to
the small components of the Dirac 4-spinor. This means that taking
the nonrelativistic limit of the baryons is a singular operation
when the pion is not soft.
In contrast, $\gamma^5\gamma^\mu$ does connect the large components
to themselves for $\mu=1,2,3$
so that, with a pseudovector coupling, we can follow the
simple leading-order large-$N_c$ prescription given earlier of just
throwing away the small components (including, \it inter alia\rm,
the $\gamma^5\gamma^0$ contribution; remember that the $1/N_c$
expansion breaks up Lorentz invariants).
Of course, in a different limit from large $N_c,$ namely
the soft-pion limit in which the pion is emitted from the on-shell
nucleon at approximately zero 4-momentum, the pseudovector and
pseudoscalar couplings are indistinguishable, provided one takes
$g_\pi^{}=g'_\pi/2M_N.$

The meaning of the words ``large-$N_c$ version of''
preceding the Lagrangian (\ref{piLagn}) is that,
as in the $\sigma N$ model of the previous section,
the coupling of the pion to the nucleon
must be supplemented by analogous couplings to all the other members of
the tower of $I=J$ baryons, and likewise for the nucleon kinetic term.
In the previous case this was an irrelevant complication:
because the zero-isospin $\sigma$ cannot induce transitions
between states in this tower, the problem diagonalizes.
In contrast, pions can and do change nucleons to $\Delta$'s, etc.

The most convenient
way to implement the gradient coupling of the pion to the $I=J$
tower of baryons is to change baryon basis to the so-called collective
coordinate basis $|A\rangle$ familiar from the Skyrme model,
with $A$ an $SU(2)$ element \onlinecite{ANW}.
 These basis elements are defined by\onlinecite{ANW,Schulman}.
%
%\eqn\eqi
\begin{equation}
\ket{A}\ =\!\!\!\!\sum_{R={1/2,\,3/2,\cdots}}(2R+1)^{1/2}
\!\!\!\!\sum_{i_z,s_z=-R,\cdots,R} (-)^{R-s_z}
D^{\scriptscriptstyle(R)}_{-s_z,i_z}(A^\dagger)
\ket{R\atop i_z\,s_z}\ ,
\label{eqi}
\end{equation}
normalizing the volume of $SU(2)$ to unity.
On the right-hand side of this equation,
the baryons are given in the usual spin-isospin
basis, e.g., a neutron of spin up and a $\Delta^+$ of spin projection $-3/2$
would be denoted as $\left|{1/2\atop-1/2,1/2}\right\rangle$ and
$\left|{3/2\atop1/2,-3/2}\right\rangle$, respectively.
In the collective coordinate language, the correct pion-baryon coupling reads
%
%\eqn\eqj
\begin{equation}
-3g_\pi\sum_{a,b=-1,0,1}\partial_b\pi^a\int_{SU(2)}dA\,
D^{\scriptscriptstyle(1)}_{ab}(A)\ket{A}\bra{A}\ .
\label{eqj}
\end{equation}
This coupling was first written down on general grounds (without
reference to soliton physics) by
Adkins, Nappi and Witten \onlinecite{ANW}, and is necessary for the
consistency of the Compton-type graphs with the overall $N_c^0$
scaling of the pion-baryon scattering amplitude in the large-$N_c$
limit, as reviewed earlier\onlinecite{GerSak,DashMan}. It has also
recently been established
 using collective coordinate quantization of the skyrmion\onlinecite{DHM}.
Despite our convenient adoption of Skyrme-model notation, we emphasize
that the coupling~(\ref{eqj}) is, in the present context, understood
to be built from explicit baryon field operators, and not solitons.

Let us verify explicitly that Eq.~(\ref{eqj}) is indeed the correct large-$N_c$
pseudovector coupling of the pion to the baryon tower. In particular,
Eq.~(\ref{eqj}) has the  following four desirable properties:
\begin{enumerate}
  \item It is invariant under isospin and angular momentum;
  \item It contains the pion-nucleon interaction shown in Eq.~(\ref{piLagn});
  \item It correctly implements the ``proportionality rule'' governing
couplings to the higher states in the $I=J$ tower.
  \item It accurately predicts the width of the $\Delta$, and furthermore,
gives widths so large for the large-$N_c$ artifacts of the model
(the baryons with $I=J\ge\textstyle{5\over2}$) that these pose no
phenomenological problems for the large-$N_c$ approach.
\end{enumerate}
\noindent
We deal with each of these assertions in turn:

1. The state $|A\rangle$ transforms as
%
%\eqn\eqk
\begin{equation}
\ket{A}\
{\stackrel{\rm\scriptscriptstyle isospin}{\longrightarrow}}\
\ket{U_I^{}A}\quad\hbox{and}\quad
\ket{A}\
{\stackrel{\rm\scriptscriptstyle ang. mom.}{\longrightarrow}}\
\ket{AU_J^\dagger}
\label{eqk}
\end{equation}
so that
%
%\eqn\eql
\begin{eqnarray}
\int_{SU(2)} dA  &\,
D^{\scriptscriptstyle(1)}_{ab}(A){\ket{A}}{\bra{A}}\ \longrightarrow\
\int_{SU(2)}dA\,
D^{\scriptscriptstyle(1)}_{ab}(A)\ket{U_I^{}AU_J^\dagger}
\bra{U_I^{}AU_J^\dagger} \nonumber\\
 = &\
\int_{SU(2)}dA\,
D^{\scriptscriptstyle(1)}_{ab}(U_I^\dagger AU_J^{})
\ket{A}\bra{A} \nonumber\\
 = &\
D^{\scriptscriptstyle(1)}_{aa'}(U_I^\dagger)
D^{\scriptscriptstyle(1)}_{bb'}(U_J^\dagger)
\int_{SU(2)}dA\, D^{\scriptscriptstyle(1)}_{a'b'}(A)\ket{A}\bra{A}\ .
\label{eql}
\end{eqnarray}
Here we have used the group invariance of the $SU(2)$ measure, $d(U_I^\dagger
AU_J^{})=dA,$ and the reality property of the
$D^{\scriptscriptstyle(1)}$ matrices. Similarly,
%
%\eqn\eqm
\begin{equation}
\partial_b\pi^a\ \longrightarrow\ \partial_{b''}\pi^{a''}
D^{\scriptscriptstyle(1)}_{a''a}(U_I^{})
D^{\scriptscriptstyle(1)}_{b''b}(U_J^{})\ .
\label{eqm}
\end{equation}
Combining these last two equations and using the composition property of the
Wigner $D$ matrices, we confirm that the coupling of Eq.~(\ref{eqj})
{\it is} invariant under isospin and angular momentum rotations.

2.  Using Eq.~(\ref{eqi}), we rewrite the coupling of Eq.~(\ref{eqj}) as
%
%\eqn\eqo
\begin{eqnarray}
-3g_\pi\sum_{a,b} \partial_b\pi^a & & \sum_{R,i_z,s_z} \sum_{R',i'_z,s'_z}
(-)^{R-s_z} (-)^{R'-s_z'}
\big[(2R+1)(2R'+1)\big]^{1/2}
\ket{R'\atop i_z'\,s_z'}\bra{R\atop i_zs_z} \nonumber\\
& &\quad\quad\quad\quad\times\
\int_{SU(2)}dA\,
D^{\scriptscriptstyle(1)}_{ba}(A^\dagger)
D^{{\scriptscriptstyle(R')}*}_{-s_z',i_z'}(A^\dagger)
D^{\scriptscriptstyle(R)}_{-s_z,i_z}(A^\dagger) \nonumber\\
&=&\
-3g_\pi\sum_{a,b}\partial_b\pi^a\sum_{R,i_z,s_z}\sum_{R',i'_z,s'_z}
(-)^{R+R'}
\langle R\,1\,i_z\,a|R'\,i_z'\rangle
\langle R'\,1\,s_z'\,b|R\,s_z\rangle \nonumber\\
& &\quad\quad\quad\quad\quad\quad\quad\quad\quad\quad\times\
\ket{R'\atop i_z'\,s_z'}\bra{R\atop i_zs_z}\ ,
\label{eqo}
\end{eqnarray}
using standard $D$-matrix integration tricks.
We now pick out the terms with $R=R'=1/2$ in this expression
in order to study specifically the pion coupling
to $\overline{N}N$. Isospin and angular momentum invariance can be
made more manifest by rewriting this subset of terms as
%
%\eqn\eqp
\begin{equation}
g_\pi\sum_{a,b}\sum_{i_z,s_z}\sum_{i'_z,s'_z}
\tau^a_{i_z'\,i_z}\sigma^b_{s_z'\,s_z}  \,
\partial_b\pi^a\ket{1/2\atop i_z'\,s_z'}\bra{1/2\atop i_zs_z}
\label{eqp}
\end{equation}
which we recognize as the nonrelativistic (or, equivalently,
in the present context, large-$N_c$) limit of the gradient coupling
$-g_\pi\partial_\mu\vec\pi\cdot\overline{N}\gamma^5\gamma^\mu\vec\tau N$.

3.  A careful reading of Ref.~\onlinecite{Muk} reveals this
criterion will be automatically satisfied due to the diagonality of the
pion-baryon coupling, Eq.~(\ref{eqj}), in the collective coordinate $A$.
It is instructive nevertheless to see how this comes about explicitly.
The baryon-antibaryon Hilbert-space operator in Eq.~(\ref{eqo}) can
be written in terms of states with good $t$-channel (exchange-channel)
quantum numbers as follows:
%
%\eqn\eqq
\begin{eqnarray}
\ket{R'\atop i_z'\,s_z'} \bra{R\atop i_zs_z}\ &=& \
\sum_{I_t,I_{tz}} \sum_{J_t,J_{tz}}
(-)^{R+i_z} (-)^{R'+s_z'}
\langle I_t I_{tz} | R' R i_z',-i_z \rangle
\langle RR's_z,-s_z' | J_t J_{tz} \rangle
\nonumber\\
& &\quad\quad\quad\quad\quad\quad\times\
\ket{{I_t\,;RR'}\atop {I_{tz}}} \bra{{J_t\,;RR'}\atop {J_{tz}}}\ ,
\label{eqq}
\end{eqnarray}
where the phases in the above are the usual cost of turning bras into
kets in $SU(2)$ \cite{RebSlan}:
 $|jm\rangle \leftrightarrow (-)^{j+m}\langle j,-m|$.
Plugging Eq.~(\ref{eqq}) into Eq.~(\ref{eqo})
 and using Clebsch-Gordan orthogonality
gives for the pion-baryon coupling:
%
%\eqn\eqr
\begin{eqnarray}
-g_\pi\sum_{I_{tz},J_{tz}} \partial_{J_{tz}}\pi^{I_{tz}}
\sum_{R,R'}& &(-)^{R+R'}
\big[(2R+1)(2R'+1)\big]^{1/2}
\ket{I_t=1\,;RR'\atop I_{tz}}
\bra{J_t=1\,;RR'\atop J_{tz}}
\label{eqr}
\end{eqnarray}
This equation correctly embodies two large-$N_c$ selection rules:
the fact that the exchanged angular momentum $J_t$ is equated to the
isospin $I_t=1$ of the pion is a specific example of the more general
$I_t=J_t$ rule \cite{Muk,Action}, whereas the square-root
proportionality
factors relating the pion's couplings to the various baryon states in the $I=J$
tower illustrate the proportionality rule \cite{Muk}.

4. The coupling (\ref{eqj}) can be used to calculate the decay width of
a baryon with spin/isospin $J$ to the next-lower state $J-1$ via the
emission of a single pion. For the case $\Delta\rightarrow N\pi$ one
calculates $\Gamma_\Delta=114\,$GeV as against a measured width of
120$\pm5\,$ MeV \onlinecite{ANW,DHM}. Pleasingly,
for the higher states, $I=J\ge
\textstyle{5\over2},$ the widths turn out to be so large that these
large-$N_c$ artifacts cannot be said to exist as particles, and
therefore, pose no phenomenological problem for the large-$N_c$
program. One finds $\Gamma_{5\over2}\approx800\,$MeV,
$\Gamma_{7\over2}\approx2600\,$MeV,
$\Gamma_{9\over2}\approx6400\,$MeV, and so forth \onlinecite{DHM}.

As before, we seek to
sum the set of exchange-type graphs of the form shown in Fig.~2.
However, \it a priori\rm, the situation is not so simple as in
the $\sigma$-only model of Sec.~II. Look again at the interesting
identity~(\ref{eqb}) for distributions, which is the key to turning
loops into trees. The $n!$ terms on the left-hand side correspond
to the $n!$ distinct ``tanglings'' in which the exchanged $\sigma$ lines are
attached in a different order to the baryon line. Because the $\sigma$
carries no spin or isospin, each tangling enters with the same
relative group-theoretic weight in Eq.~(\ref{eqb}), and the identity
goes through as written (so, too, for photon exchange).
 In contrast, $\pi,$ $\rho$ and  $\omega$ mesons, etc., carry non-trivial
isospin and/or spin, and the $n!$ tanglings would not be expected to
occur with the same group-theoretic factors.
 (Pauli spin/isospin matrices do not
commute.) Specifically, one expects a different product of $n$ spin
and $n$ isospin Clebsch-Gordan factors weighting each term on the
left-hand side of Eq.~(\ref{eqb}), and destroying the identity.
Nevertheless, acting together, the $I_t=J_t$ and proportionality rules
assure that, to \it leading \rm order in $1/N_c,$ these $n!$ group-theoretic
factors  are indeed equal, once the intermediate baryon legs are
summed over all allowed $I=J$ states. Therefore,
the identity~(\ref{eqb}), derived for $\sigma$ (or photon) exchange, applies
as well to the exchange of these non-trivial mesons.
This theorem is proved in  Ref.~\onlinecite{AM}, using elementary
properties of 6$j$ symbols. However, there is an easier way to
see this, which is to work directly in the $|A\rangle$ basis.
 So, look again at Fig.~5, and understand the
 baryon line to mean, not a nucleon or
a $\Delta$ or any specific member of the $I=J$ tower (which can  change
identity at each pion interaction vertex), but rather a baryon state
$|A\rangle$ sharp in the $SU(2)$ collective coordinate $A,$ which is
\it preserved \rm at each vertex,
due to the diagonality in $A$ of the coupling (\ref{eqj}).
Initial and final nucleon, $\Delta,$
etc., states can be projected out at the very end of the calculation
using standard group-theoretic techniques
borrowed from the Skyrme model [i.e., inverting Eq.~(\ref{eqi})].
At earlier stages, however, we can use the full machinery of Sec.~II to
turn loops into trees with impunity.

Therefore, once again, the graphs of Fig.~5 can be summed
following a two-stage program. In the first stage, one solves a static
non-linear equation for $\vec\pi_{\rm cl}(A)$ (noting that the
classical pion field depends on the $SU(2)$ collective coordinate
$A$). Isospin covariance trivially
relates this quantity to $\vec\pi_{\rm cl}(A=1)$, henceforth called
just $\vec\pi_{\rm cl}.$
Using $D^{\scriptscriptstyle(1)}_{ab}(A=1)=\delta_{ab},$
one obtains the Euler-Lagrange equation
%
%\eqn\eqs
\begin{equation}
-\nabla^2\pi_{\rm cl}^a+m_\pi^2\pi_{\rm cl}^a
+{1\over6}\lambda\pi_{\rm cl}^a\vec\pi^{}_{\rm cl}\cdot\vec\pi^{}_{\rm cl}
-3g_\pi\partial_a \delta^3({\bf x})\ =\ 0\ .
\label{eqs}
\end{equation}
This equation is solved by smearing the $\delta$-function source to
a Gaussian as in Eq.~(\ref{gaussian}), and by
assuming a hedgehog Ansatz for the classical pion field
(anticipating the resemblance to the Skyrme model):
\begin{equation}
  \pi _{\rm cl}^a({\bf x})={\hat {\bf r}}^a F(r) \ .
  \label{HHA}
\end{equation}
Equation (\ref{eqs}) then becomes an ODE for the classical pion profile $F(r)$:
\begin{eqnarray}
F''  + {2 \over r}F' \
  & - &\ ({2 \over r^2}+m_\pi^2) \; F  - {\lambda\over6} F^3
  = \ -{6g_\pi r\over a_N^5\pi^{3/2}}\, \exp(-r^2 / a_N^2)  \ ,
  \label{Fonlyeqn}
\end{eqnarray}
subject to the boundary conditions that
$F(r)$ be regular near $r=0$ and bounded as $r \to \infty$,
\begin{eqnarray}
  F(r) &=&   \; Br + {\cal O}(r^3) \ {\rm near} \ r = 0; \nonumber \\
  F(r) &\to& \; C\exp(-m_\pi r)/r \ {\rm as} \  r \to \infty \ .
  \label{piBCs}
\end{eqnarray}
$B$ and $C$ are scale parameters that are initially unknown to us but are
fixed implicitly by the non-linearity of Eq.~(\ref{Fonlyeqn}).
This is another two-boundary-value problem, which can be
numerically solved as before (see Fig.~11, and Appendix A).

In the second stage, one solves the linearized time-dependent equation
for $\pi_{\rm qu}$ propagating in the background of $\pi_{\rm cl}(A).$
Again, isospin invariance trivially relates this process to the propagation
of $\pi_{\rm qu}$ in the background of $\pi_{\rm cl}(A=1),$ the latter
quantity being given by Eqs.~(\ref{HHA}) and (\ref{Fonlyeqn}).
Initial and final nucleons or $\Delta$'s are then
projected  from the hedgehog by inverting Eq.~(\ref{eqi}),
using the orthogonality over $SU(2)$ of Wigner $D$-matrices.
Finally, the initial and final pion-baryon
systems are combined into states of good total isospin and angular
momentum in the usual fashion to give the partial-wave $S$-matrix
for $\pi N\rightarrow\pi N$,  $\pi N\rightarrow\pi\Delta$, etc.

Fortunately, this cumbersome (if straightforward)
sequence of group-theoretic steps can
be circumvented, once one realizes that they are \it identical \rm to
the procedure followed in the Skyrme model \cite{HEHW,Sig,MandK,MandP}.
Rather than
``reinventing the wheel'' one can therefore carry over intact the
machinery of Refs.~\onlinecite{HEHW,Sig,MandK,MandP} of
$K$-spin decomposition and 6$j$ symbols.
We postpone the explicit review of this formalism
to Sec.~\ref{physS}, in which we complete the analysis of
 the richer model containing both pions and $\sigma$ mesons.

Unfortunately, the pion-only model discussed in this Section
is inherently uninteresting  phenomenologically. Because of $G$-parity,
the pion-pion interactions can only come from even powers of $\vec{\pi}(x)$,
which means that the potentials entering into the coupled Schr\"odinger-like
scattering equations are strictly repulsive.
[They are proportional to $\lambda F^2(r)$.]
As a result, all $\pi N$ phase shifts exhibit repulsive behavior (i.e.,
clockwise motion in the Argand plots with increasing energy).
Thus there is no possibility for $\pi N$ resonances in such a model.
We need {\it something} like the $\sigma$ meson to provide a range
of attraction between $\pi$'s and $N$'s.

\section{Defining the $\sigma$-$\pi$ model}
   \label{sec:sigpimodel}

In view of the two models discussed  in the two previous Sections,
 one might have some hope that a model combining $\sigma$ and $\pi$
mesons would provide a more promising (if still crude)
description of pion-nucleon interactions.
In this model the $\sigma$ meson will be taken as an ``elementary''
 field, along
with the three $\pi$ fields.  Indeed, in the large-$N_c$ limit, the $\sigma$,
if such a state exists, is necessarily a stable particle, as
the decay amplitude to two pions is suppressed by  $1 / \sqrt N_c$.

For guidance in constructing our large-$N_c$ model of pions and $\sigma$
mesons, and selecting reasonable values of the coupling constants,
we recall the linear $\sigma$-model of Gell-Mann and Levy:\cite{linsigmod}
\begin{eqnarray}
{\cal L}  &=&
    {1 \over 2} \partial_\mu \sigma' \partial^\mu  \sigma'
    + {1 \over 2} \partial_\mu \vec\pi \cdot  \partial^\mu  \vec\pi
    - {\lambda\over 4} ({\sigma'}^2
    + \vec\pi\cdot\vec\pi - a^2 )^2
    + \alpha  \sigma' \nonumber \\
& & \quad
    - g \sigma' \overline{N}N
    - g \vec\pi \cdot \overline{N} i\gamma^5\vec\tau  N
+\overline{N}i\gamma\cdot\partial N\ .
\label{sigpiLagn}
\end{eqnarray}
In this well-known model, the nucleon and $\sigma$ get their masses through
dynamical symmetry breaking, the $\sigma$ vacuum
expectation value $v$ being $g^{-1}M_N,$ and
chiral symmetry emerges in the limit $\alpha \rightarrow 0$.
It is convenient to redefine the $\sigma$ field by subtracting the VEV,
%
%\eqn\eqt
\begin{equation}
   \sigma'(x)\ = \ v+\sigma(x) \ .
\label{eqt}
\end{equation}
By substituting for $\sigma'$ and expanding,
the four coupling constants $\{g,\lambda,a,\alpha\}$ can be traded for
the more physical set of parameters, $\{g, M_N, m_\pi, m_\sigma\}$, using
\begin{eqnarray}
  \lambda  = {g^2 \over 2M_N^2} (m_\sigma^2 - m_\pi ^2)\ , \quad\quad
  \alpha  = {{m_\pi}^2 M_N \over g} \ , \quad\quad
  a^2  = {M_N^2 \over g^2}
      {(m_\sigma^2 - 3m_\pi^2) \over (m_\sigma^2 - m_\pi^2)}
  \label{params}
\end{eqnarray}
In this paper we will take
\begin{equation}
 g=13.6\ ,\quad
 M_N = 5.0 \ {\rm  fm}^{-1} \ , \quad m_\pi = 0.7 \ {\rm  fm}^{-1} \ ,
  \quad {\rm and } \quad m_\sigma = 5.0 \ {\rm  fm}^{-1} \ .
\label{params1}
\end{equation}
This choice for the nucleon mass roughly averages the actual $N$ and $\Delta$
masses, while the $\sigma$ meson here could be
identified with the $f_0(975)$ meson for concreteness.
The value of $g$ is the measured pion-nucleon pseudoscalar coupling constant.
With these values, the non-linear
self-interaction strength has a large value, $\lambda \approx 91$.

For a large-$N_c$ treatment, the Gell-Mann-Levy model needs to be
modified in the following two ways.
First, as discussed in Sec.~\ref{pionlymodel},
the pseudoscalar $\pi N$ coupling is inappropriate,
and should be replaced by pseudovector coupling as in Eq.~(\ref{piLagn}),
with $g_\pi=g/(2M_N) = 1.42$ fm.
Unfortunately, with this replacement chiral symmetry is lost, even
for $\alpha=0.$ However, as stated in the introduction,
our purpose in this paper is to explore the large-$N_c$ approach in
a multi-channel model, not to present a fully realistic effective
Lagrangian of the low-lying hadrons, which would require not only
approximate chiral symmetry but also the incorporation of vector mesons.
(To look at the bright side, the fact that we are sacrificing chiral
symmetry re-emphasizes the point that our large-$N_c$ techniques
have nothing to do with the chiral limit.)

Second, the meson couplings to the nucleon must be augmented by
suitable couplings to the entire $I=J$ baryon tower (and likewise
for the nucleon kinetic energy). The prescription
for doing so is Eq.~(\ref{eqj}) for the pion. It is easy to check that the
analogous prescription for the $\sigma$ is given simply by
%
%\eqn\equ
\begin{equation}
-g\sigma\overline{N}N\ \longrightarrow\
-g\sigma\int_{SU(2)}dA\,\ket{A}\bra{A}\ .
\label{equ}
\end{equation}
%
%{\bf (MOVE THIS COMMENT:  Note
%the arbitrariness of choosing $M_N$ to be its physical value in a model in
%which it is also assumed to be of order $N_c$,  i.e., very massive.)}
%The
% remaining parameters in $\cal L$ to be fixed are the coupling constants for
%the meson-nucleon interactions.
%Using the equivalence theorem relating pseudo-vector and pseudo-scalar
%coupling, we again emulate the linear $\sigma$-model symmetry by choosing
%
%\begin{equation}
%\matrix{
%  g_\pi \, =\, {g / 2M_N}\, {\rm ,}\quad \,
%  g_\sigma \, =\, g\, {\rm ,} \cr }
%  \label{params2}
%\end{equation}
%
%where $g = g_{\pi NN} = 13.6$ is the usual pion-nucleon pseudoscalar coupling
%constant.
%With this value for $G$ and the masses chosen as above, $\lambda$ = 90.66,
%which is a {\em large} value from the standpoint of the non-linearity of the
%resulting equations for the classical field solutions.
%
%We also introduce one more parameter, the size of the nucleon, $a_N$, when we
%freeze out the nucleon degree of freedom.
%In doing so we replace the nucleon by a smeared, static source, i.e.,
%
%\begin{equation}
%  \overline{N}N \to {1 \over {(a_N \sqrt{\pi})^3}} \exp(-r^2/a_N^2)
%\end{equation}
%
%{\bf [Again, how to present the replacement of the PV term?]}

As previously, we solve for the classical meson fields, for
the reference choice of collective coordinate $A=1$, by means of
 a hedgehog ansatz:
\begin{equation}
  \pi _{\rm cl}^a({{\bf x}})={\hat {\bf r}}^a F(r) \ , \quad
  \sigma_{\rm cl}({\bf x}) = G(r) \ .
  \label{HHAsigpi}
\end{equation}
Smearing out the $\delta$-function baryon source as in Eq.~(\ref{gaussian}), we
find coupled non-linear Euler-Lagrange ODE's for $F$ and $G$:
\begin{mathletters}
\begin{eqnarray}
{d^2  \over {dr}^2 }F  + {2 \over r}{d \over dr}F
  & - & \left({{2 \over r^2 }+m_\pi ^2}\right)\, F
  - \lambda \left[{F^3 +FG^2 +2v FG}\right] \nonumber \\
  & = & - {3g r\over M_N^{}
a_N^5\pi^{3/2}} \exp\left({-{r^2  \big/ a_N^2}}\right) \\
{d^2  \over {dr}^2 }G + {2 \over r}{d \over dr}G
  & - & m_\sigma ^2\, G
  - \lambda \left[{G^3 +F^2 G+3v G^2}\right] \nonumber \\
  & = &  {g\over(a_N\sqrt{\pi})^3} \exp\left({-{r^2  \big/ a_N^2}}\right)
  + \, \lambda v F^2\ .
  \label{Geqn}
\end{eqnarray}
  \label{FGeqns}
\end{mathletters}
We will generally set the nucleon size parameter
$a_N$ = 0.52 fm, but we will also consider the
dependence of our results on $a_N$ in Sec.~VI(C) below.
The boundary conditions are that
$F$ and $G$ must be regular at the origin and exponentially decaying (rather
than growing) at infinity.
The classical pion profile $F(r)$ falls off like $\exp(-m_\pi r)/r$
at large distances. On the other hand, $G(r)$ falls off not like
$\exp(-m_\sigma r)/r$ as one might naively expect,
but rather like
$\exp(-2m_\pi r)/r^2$ due to the $F^2$ source term on the
right-hand side of Eq.~(\ref{Geqn}), and the fact that $2m_\pi < m_\sigma$.
Details of our numerical ``shoot and match''
procedure for solving Eq.~(\ref{FGeqns}) can be found in Appendix A.

The solution for $F(r)$ and $G(r)$ is shown in Fig.~12.
Note that $G(r)$ is negative with respect to $F(r)$ and $v$.
It is this relative sign that leads to the attractive $\pi N$
interaction found in this model.

\section{pion-hedgehog scattering}
 \label{QScat}

Having solved for the classical pion and $\sigma$ fields, we
turn to the  small-fluctuations
problem of meson-baryon scattering. As in the
Skyrme model\onlinecite{HEHW,Sig,MandK,MandP}, one first
solves for meson-{\it hedgehog} scattering, and subsequently one
folds in some group theory (6$j$ symbols) to obtain meson-{\it nucleon}
scattering. The meson-hedgehog $S$-matrix is the topic of
this Section, while the meson-nucleon $S$-matrix is the subject of
 Section VI to follow.

We return to the $\sigma$-$\pi$
Lagrangian, Eq.~(\ref{sigpiLagn}) as modified subsequently in the
text in the manner suggested by  large-$N_c$. Consider fluctuations of the
meson fields about their classical solutions,
\begin{equation}
   \pi^a(x) \to {\bf \hat{r}}^a F(r) + \pi_{\rm qu}^a(x) \quad ,\quad
   \sigma(x) \to G(r) + \sigma_{\rm qu}(x) \ .
   \label{qflucs}
\end{equation}
Since $F$ and $G$ satisfy the Euler-Lagrange equations,
 terms linear in the fluctuating fields vanish.
The quadratic terms then lead to linear
equations of motion for  $\pi_{\rm qu}^a(x)$ and $\sigma_{\rm qu}(x)$.
Higher-order nonlinearities in the meson fields are subleading
in $1/N_c$, as previously noted.

We will work out the partial-wave scattering
 amplitudes factoring out a uniform
 time-dependence  $\exp(- i \omega t)$ from all the fluctuating fields.
For the  $\sigma$ this involves the usual expansion in
spherical harmonics,
\begin{equation}
   \sigma_{\rm qu}(\omega,{\bf x}) = \sum_{K,K_z} \phi_{KK_z}
(\omega,r) \; Y_{KK_z}(\hat{\bf x})
   \label{sigmaPW}
\end{equation}
For the pions the decomposition is slightly more complicated
\cite{HEHW,Sig,MandK,MandP}.
The conserved quantum numbers are not isospin and total angular
momentum but the so-called ``grand spin,'' $\vec{K} = \vec{I}+\vec{J}$.
Since pions are spinless, $\vec J$ is just $\vec L,$
the orbital angular momentum.
Thus the appropriate partial wave analysis for pions involves an expansion in
terms of {\em vector} spherical harmonics,
\begin{equation}
   \vec{\pi}_{\rm qu}
(\omega,{\bf x}) = \sum_{K,K_z,L} \psi_{KK_zL}(\omega,r) \;
     \vec{\cal Y}^{\ L}_{KK_z}(\hat{\bf x}) \ ,
   \label{piPW}
\end{equation}
where $L$ runs over values $K-1$, $K$, and $K+1$.

For each value of $K$, the equations for the four radial
wavefunctions $\phi_K,$ $\psi_{K,K},$ and $\psi_{K,K\pm1}$
 might be expected to form a $4 \times 4$ coupled system,
\footnote{\divide\baselineskip by 2
  From now on we drop the $K_z$ label on $\phi$ and $\psi$ since the ensuing
  equations are independent of $K_z$.}
but parity uncouples $\psi_{K,K}$ from the other three. It obeys
\begin{equation}
  {d^2 \over {dr^2}}\psi_{K,K} +
    {2 \over r}{d \over dr}\psi _{K,K} +
    \left[\,q_\pi^2 - {K(K+1) \over r^2 } - V_\pi(r)\,\right]
 \psi_{K,K} = 0 \ ,
  \label{uncpld}
\end{equation}
where
\begin{equation}
q_\pi^2 = \omega^2 - m_\pi^2
\quad\hbox{and}\quad
  V_\pi(r) = \lambda [F^2(r) + G(r)(2 v + G(r))] \ .
  \label{Vpi}
\end{equation}

\def\aplus{\sqrt{K+1\over2K+1}}
\def\aminus{\sqrt{K\over2K+1}}
The remaining
$3 \times 3$ coupled system of equations\footnote{
\divide\baselineskip by 2
For the special
case $K=0$ this is a 2$\times$2 as $\psi_{0,-1}$ does not exist.} is
best expressed in matrix form. Assembling $\psi_{K,K\pm1}$ and
$\phi_K$ into the column vector
\begin{equation}
  \Psi_K(r) = \left( \begin{array}{c}
                     \psi_{K,K-1}(r)\\
                     \psi_{K,K+1}(r)\\
                     \phi_K(r)
                     \end{array} \right)  \ ,
  \label{defPsifirst}
\end{equation}
we find\footnote{
\divide\baselineskip by 2
In so doing we are greatly assisted by the vector
spherical harmonic identities, Eq.~(10), in Ref.~\onlinecite{MandK}.
Note a typo there: $K$ in the numerator of the square-root
in the middle line of Eq.~(10) should instead be $K+1$.}
\begin{equation}
  {d^2 \over {dr}^2}\Psi_K  +  {2 \over r}{d \over dr}\Psi_K +
\left[{\sf Q}_K-{\sf V}_K\right]\cdot\Psi_K=0\ .
  \label{cpld}
\end{equation}
Here ${\sf Q}_K$ is the diagonal matrix
\begin{equation}
{\sf Q}_K={\rm diag}
  \left(q_\pi^2-{(K-1)K\over r^2},\ \ q_\pi^2-{(K+1)(K+2)\over r^2},\ \
  q_\sigma^2-{K(K+1)\over r^2}\right)\, ,
  \label{QKdef}
\end{equation}
and ${\sf V}_K$ is the symmetric potential energy matrix
\begin{mathletters}
\begin{eqnarray}
  {\sf V}_{11} \;&=&\;
       V_\pi(r) + 2 \lambda F^2(r) \left({K \over 2K+1}\right) \\
  {\sf V}_{12} \;&=&\;  - 2 \lambda F^2(r) \left({
       \sqrt{K(K+1)} \over 2K+1}\right) \\
  {\sf V}_{13} \;&=&\;
       2 \lambda F(r)( v + G(r)) \left({K \over 2K+1}\right)^{1/2} \\
  {\sf V}_{22} \;&=&\;
       V_\pi(r) + 2 \lambda F^2(r) \left({K+1 \over 2K+1}\right) \\
  {\sf V}_{23} \;&=&\;
      -2 \lambda F(r)( v + G(r)) \left({K+1 \over 2K+1}\right)^{1/2} \\
  {\sf V}_{33} \;&=&\;  V_\sigma(r)  \ ,
  \label{defV}
\end{eqnarray}
\end{mathletters}
where we have defined
\begin{equation}
  q_\sigma^2 = \omega^2 - m_\sigma^2 \quad\hbox{and}\quad
  V_\sigma(r) = \lambda [F^2(r) + 3 G(r)(2 v + G(r))] \ .
  \label{Vsigma}
\end{equation}
Note that $q_\sigma^2$ can be positive or negative, depending on whether the
energy $\omega$ is above or below the $\sigma$ threshold.

The ``diagonal'' potentials $V_\pi$ and $V_\sigma$ are plotted in Fig.~13.
They are repulsive at short distances and attractive at intermediate range.
The factor of three in the definition of
$V_\sigma$ makes it about three times more repulsive and
attractive than $V_\pi$.
Note that the vertical scale is in inverse fermis; these are potential wells of
depths about 6 and 2 GeV, respectively, which means there is substantial
attraction in both the $\sigma N$ and $\pi N$ systems.
Also shown in Fig.~13 are the off-diagonal transition potentials
${\sf V}_{12}$ and ${\sf V}_{13}$ (but without $K$-dependent factors)
which  are comparable in size to the  diagonal potentials.

Numerically, the uncoupled
equation (\ref{uncpld}) is readily solved using the Runge-Kutta
technique employed in Secs.~II and III. This method also works for
the coupled equations (\ref{cpld}), but only
\it above \rm the $\sigma$-threshold,  $\omega > m_\sigma$. The
problem below threshold  is to ensure that the
$\sigma$ wavefunction remains exponentially decaying,
\begin{equation}
  \phi_K(r) \to \exp(- \kappa r)/r \ ,\quad
\kappa = (m_\sigma^2 - \omega^2)^{1/2}\ .
  \label{sigdecay}
\end{equation}
In our experience, numerical noise in the Runge-Kutta integration
invariably induces
exponential blow-up: $\phi_K(r) \to \exp(+ \kappa r)/r$. We emphasize
that even below threshold the $\sigma$ cannot be neglected as
it causes substantial attraction in the $\pi N$ channel.
(Recall that the ``box diagram'' for $\pi N \to \sigma N \to \pi N$,
Fig.~2c, is attractive.)

A numerically more robust approach that works both above and
below the  $\sigma$-threshold is to convert
  Eq.~(\ref{cpld}) into a set of
coupled Fredholm integral equations of the second kind,
\begin{equation}
  \Psi^{(i)}_K(r) =  {\cal J}^{(i)}_K(r)
 + \int {\sf G}_K(r,r') {\sf V}_K(r') \Psi^{(i)}_K(r') \, dr' \ ,
  \label{IntEqn}
\end{equation}
where the index $i$  labels the linearly independent choices of
inhomogenous driving terms.
Above the $\sigma$ threshold, $i$ runs over 1,2,3 and the inhomogeneous
terms are
%
%\begin{mathletters}
\begin{eqnarray}
  {\cal J}^{(1)}(r) = \left( \begin{array}{c}
             \hat{\jmath}_{K-1}(q_\pi r) \\ 0 \\ 0
             \end{array} \right)  \ , \quad
  {\cal J}^{(2)}(r) = \left( \begin{array}{c}
             0 \\ \hat{\jmath}_{K+1}(q_\pi r) \\ 0
             \end{array} \right)  \ , \quad
  {\cal J}^{(3)}(r) = \left( \begin{array}{c}
            0 \\ 0 \\ \hat{\jmath}_{K}(q_\sigma r)
             \end{array} \right)     \ .
  \label{defJ}
\end{eqnarray}
%\end{mathletters}
%
Below threshold, only the first two of these should be kept.
The multi-channel Green's function  ${\sf G}_K$ is the diagonal matrix
\begin{eqnarray}
  {\sf G}_{11}(r,r') =
- &&{1\over q_\pi} \hat{\jmath}_{K-1}(q_\pi r_<) \hat{n}_{K-1}(q_\pi r_>)\ ,
     \nonumber \\
  {\sf G}_{22}(r,r') =
- &&{1\over q_\pi} \hat{\jmath}_{K+1}(q_\pi r_<) \hat{n}_{K+1}(q_\pi r_>)\ ,
     \\
  \label{defG}
  {\sf G}_{33}(r,r') =
- &&{2\over{\pi\kappa}}\hat{\imath}_{K}(\kappa r_<)\hat{k}_{K}(\kappa r_>)\ ,
     \ {\rm below\ threshold} \ , \nonumber \\
  {\sf G}_{33}(r,r') =
- &&{1\over q_\sigma} \hat{\jmath}_{K}(q_\sigma r_<) \hat{n}_{K}(q_\sigma r_>)
     \ , \ {\rm above\ threshold}    \ .  \nonumber
\end{eqnarray}
%
%\begin{eqnarray}
%  {\sf G}_{11}(r,r') =
%
%
%- &&{1\over q_\pi} \hat{\jmath}_{K-1}(q_\pi r_<) \hat{n}_{K-1}(q_\pi
%r_>)\ ,
%     \nonumber \\
%  {\sf G}_{22}(r,r') =
%
%
%- &&{1\over q_\pi} \hat{\jmath}_{K+1}(q_\pi r_<) \hat{n}_{K+1}(q_\pi
%r_>)\ ,
%     \\
%  \label{defG}
%  {\sf G}_{33}(r,r') =
%
%
%- &&{2\over{\pi\kappa}}\hat{\imath}_{K}(\kappa r_<)\hat{k}_{K}(\kappa
%r_>)\ ,
%
%     \ {\rm below\ threshold} \ , \nonumber \\
%  {\sf G}_{33}(r,r') =
%
%- &&{1\over q_\sigma} \hat{\jmath}_{K}(q_\sigma r_<)
%\hat{n}_{K}(q_\sigma r_>)
%     \ , \ {\rm above\ threshold}    \ .  \nonumber
%\end{eqnarray}
%
where $\hat{\jmath}_l$, $\hat{n_l}$ are spherical Ricatti-Bessel functions
\cite{Taylor} and  $\hat{\imath}_l$, $\hat{k_l}$ are modified spherical
Ricatti-Bessel functions \cite{AandS}, regular at the origin and exponentially
decaying, respectively.
By design, the multi-channel Green's function assures regularity of
the wave functions at the origin {\em and} the asymptotic exponential
fall-off of the $\phi_K$ below the $\sigma$ threshold.
Note that $G_{33}$ is continuous through the threshold.

The $S$-matrix for the uncoupled pion scattering, Eq.~(\ref{uncpld}),
 will be denoted here as the single-subscript quantity $S_K$,
where the orbital angular momentum quantum numbers $L=L'=K$ are suppressed.
It is derived from the asymptotic analysis of the wavefunction in the
usual way. The  corresponding phase-shift $\delta_K$, defined as
\begin{equation}
S_K=e^{2i\delta_K}\ ,
\end{equation}
is plotted against pion momentum $k$ in Fig.~14 for $K\le5$. For each
$K$ the corresponding phase shift is attractive, if numerically small
apart from the case $K=1,$ and comparatively much less significant
than in the Skyrme model ($cf$. Fig.~1, Ref.~\onlinecite{MandK}).
As always in scattering problems,
 the centrifugal barrier term in the scattering equations delays
the onset of the rise in the phase-shift for the higher-$L$ partial waves.

The coupled-channels $3 \times 3$ (above threshold)
or $2 \times 2$ (below threshold) part of the $S$-matrix
 will be denoted ${\sf S}^K_{ij}$, $i,j = 1,2$ and/or 3,
according to $L= K-1,K+1,$ and/or $K$. It is obtained
 as follows. First, the ${\sf K}^K$-matrix is formed according to
\begin{equation}
  {\sf K}^K_{ij} =  -(1/q_j) \int dr\, \hat{\jmath}_L(q_j r)
     [{\sf V}(r) \Psi^{(i)}(r)]_{j} \ ,
  \label{Kmat}
\end{equation}
where $L = K-1,K+1$ and/or $K$ for $j = 1,2$ and/or 3, respectively,
and also $q_1=q_2=q_\pi,$ $q_3=q_\sigma.$ From
the ${\sf K}^K$-matrix, the $S$-matrix is formed in the
usual way,
\begin{equation}
  {\sf S}^K_{ij} = (q_j / q_i)^{1/2}
     [(1 - i{\sf K}^K) (1 + i{\sf K}^K)^{-1}]_{ij} \ ,
  \label{KtoS}
\end{equation}
where for an explanation of the square-root flux factors (needed only
for multichannel scattering) we refer the
reader to Ref.~\onlinecite{TaylorFactors}. Time-reversal
invariance implies ${\sf S}^K=\big({\sf S}^K\big)^T,$ which we have
found to be a stringent check on our numerics. We will parametrize
${\sf S}^K_{ij}$ as $\eta^K_{ij}\exp 2i\delta^K_{ij}$ subject to this
symmetry property as well as to unitarity.

The phase-shifts corresponding to the specific $S$-matrix elements
${\sf S}^K_{11}$ and ${\sf S}^K_{22}$
 are plotted in Figs.~15-16. Recall that, with our notation,
these are the $S$-matrix elements that describe pion-baryon scattering
(no `in' or `out' $\sigma$'s, only intermediate $\sigma$'s)
in which the orbital angular momentum quantum number is preserved
$(L=L'),$ as opposed to changing up or down by two units (as it
can for $\pi N\rightarrow\pi\Delta$).
The bulk of the attraction in the present model, due primarily
to the intermediate $\sigma$-meson states, shows up in
 the phase-shifts of Fig.~16, with $L=L'=K+1$.
Here one sees resonances (phase shifts rising rapidly
through 90 degrees) in each partial wave. In the $L=L'=K-1$
partial waves (Fig.~15), one also sees attraction, although not so strong as
to produce resonances. The surprise here
 is in the channel $K=1,$ $L=L'=0$, which reveals the
presence of a \it bound state \rm (Levinson's theorem).
Once one folds in the appropriate group theory in the following Section
to project the hedgehog onto physical baryons, the existence of
such a bound state manifests itself as a parity conjugate to the nucleon.
This feature is, unfortunately, \it not \rm found
in Nature, nor in the Skyrme model, and is an
unwanted, unphysical artifact of the present strongly-coupled
$\sigma$-$\pi$ model.

On the other hand, an \it improvement \rm over the Skyrme model is the
fact that all these phase-shifts (as well as those not plotted) eventually
return to zero for sufficiently high energies. In contrast, in the
Skyrme model they apparently grow without bound,
 eventually violating the unitarity constraints
 of quantum field theory, although admittedly
at energies where several key approximations made
 in Refs.~\onlinecite{HEHW,Sig,MandK,MandP}, such as the neglect
of skyrmion recoil, are clearly unwarranted.

Another point to note about Fig.~15 are the cusp effects due to the
opening of the $\sigma$ threshold at $\omega = 5 \ {\rm fm}^{-1}$.
This is most apparent in the $K=1, L=0$ phase shift, but the effect is
present in the higher partial waves as well.

\section{$\pi N$ ELASTIC SCATTERING}
  \label{physS}

   \subsection{Group-theoretics for meson-nucleon scattering}

In the previous Section we derived an $S$-matrix for the scattering
of pions and $\sigma$'s off hedgehogs. The scattering information
is encoded in partial-wave amplitudes we called $S_K$ and
${\sf S}^K_{ij}$ where ${\sf S}^K$ is a $2\times2$ matrix below the
$\sigma$ threshold and a $3\times3$ matrix above it (except when $K=0$
in which case ${\sf S}^K$ is  $1\times1$ or  $2\times2$,
respectively). The integer index $K$ labels the vectorial sum
of the incoming or outgoing meson's isospin and angular momentum. $K$
is conserved when the meson scatters off an object with hedgehog
symmetry, in the same way that orbital angular momentum $L$ is
conserved in scattering from a spherically symmetric potential.

Of course, what we are really interested is scattering, not from
a hedgehog, but rather from a nucleon or $\Delta.$ The relationship
between the two problems, ``physical scattering'' versus ``hedgehog
scattering,'' is contained in the following group-theoretic
expression\onlinecite{HEHW,Sig,MandK,MandP,Karliner,ninj}:
\begin{eqnarray}
  S_{LL'\, RR'\, I_{\rm tot}J_{\rm tot}}(\omega) \, = \,
    \sum_K\,  S_{KLL'}(\omega) \cdot
    &&(-)^{R'-R} [(2R+1)(2R'+1)]^{1/2} (2K+1) \nonumber\\
    &&\times \left\{\matrix{K&I_{\rm tot}&J_{\rm tot}\cr R&L&I\cr }\right\}
      \left\{\matrix{K&I_{\rm tot}&J_{\rm tot}\cr R' &L' &I'\cr }\right\}\ .
  \label{Sphysical}
\end{eqnarray}
Here $\omega$ is the meson energy in the baryon rest frame (baryon recoil
being subleading in $1/N_c$), $L$
($L'$) is the initial (final) orbital angular momentum, $I$ ($I'$)
is the isospin of the incoming (outgoing) meson, and $R$ ($R'$) is
the spin/isospin of the initial (final) $I=J$ baryon (e.g., $R=1/2$
for a nucleon, $R=3/2$ for a $\Delta$, etc.). For physical scattering,
$K$ is no longer conserved;  it is just a dummy of summation,
constrained by the triangle inequalities implicit
in the 6$j$ symbols.\footnote{
\divide\baselineskip by 2
Note: if either the incoming or the
outgoing meson is a $\sigma,$ then the associated 6$j$ symbol has a zero
in it and collapses to a product of Kronecker $\delta$'s. Conversely, the
generalization of this expression to mesons that carry both isospin and
spin, such as $\rho$'s, involves 9$j$ symbols, and is given in
Ref.~\onlinecite{ninj}.}
Instead, the conserved quantities are, as they must be, the total meson+baryon
isospin and angular momentum, $I_{\rm tot}$ and $J_{\rm tot}$.
The $S$-matrix element on the left-hand side is a physical
partial-wave amplitude that can be compared directly with experiment.
The ``reduced $S$-matrix'' under the summation is a meson-hedgehog
amplitude, in slightly different notation than that of the previous
Section. The relation between the two notations is: when the incoming
and outgoing mesons are each pions, then $S_{KKK}=S_K$,
$S_{K,K-1,K-1}={\sf S}^K_{11}$, $S_{K,K+1,K+1}={\sf S}^K_{22}$,
$S_{K,K-1,K+1}=S_{K,K+1,K-1}={\sf S}^K_{12}$; when they are both $\sigma$'s
then $S_{KKK}={\sf S}^K_{33}$; and when the incoming meson is a pion and the
outgoing meson is a $\sigma$, then $S_{K,K-1,K}={\sf S}^K_{13}$
and $S_{K,K+1,K}={\sf S}^K_{23}$; with all other elements vanishing.

   \subsection{The Big-Small-Small-Big pattern}

For the remainder of this paper we specialize to the elastic case
$\pi N\rightarrow\pi N.$ For each value of $L=L',$ there are then four
\it a priori \rm independent partial wave amplitudes, traditionally denoted
$L^{}_{2I_{\rm tot},2J_{\rm tot}}$. For example, in the case of
$F$-wave scattering ($L=3$) the four physical amplitudes are $F_{15},$
$F_{17},$ $F_{35},$ and $F_{37}.$ But to leading-order in
large-$N_c,$ only two out of these four are independent. One can,
for instance, solve for the two isospin-$3/2$ amplitudes as energy-independent
linear combinations of the two isospin-$1/2$ amplitudes\onlinecite{HEHW,MandP};
this is an example of the $I_t=J_t$ rule\onlinecite{Muk,Action}. This holds
in the Skyrme model, and because the group-theoretic expression
  (\ref{Sphysical})
is the same, necessarily in the present $\sigma$-$\pi$ model as well.
These relations are reasonably well obeyed by the experimental $\pi N$
partial-wave data\onlinecite{MandP}, and are model-independent
tests of large $N_c.$

Another interesting  fact about the experimental data (see Fig.~4
\it ff.\rm,
Ref.~\onlinecite{MandP}):  If for each $L$ one juxtaposes the
four amplitudes in the above order, namely $L_{1,2L-1}$, $L_{1,2L+1}$,
$L_{3,2L-1}$ and $L_{3,2L+1}$, then they reveal a striking pattern
termed the ``Big-Small-Small-Big'' pattern. Namely, the  outer two
amplitudes, $L_{1,2L-1}$
and $L_{3,2L+1}$, are characterized by relatively large
excursions of the $S$-matrix element through the Argand circle, while
the inner two amplitudes, $L_{1,2L+1}$ and $L_{3,2L-1}$, show relatively
much less motion. The Big-Small-Small-Big pattern is the
single most consistent pattern characterizing the partial-wave $S$-matrix
as a whole (the only clear exception to it being the $D_{35}$).

Reproducing the Big-Small-Small-Big pattern is one of the noteworthy
successes of the Skyrme model\onlinecite{MandK}. It is equally
well reproduced by the present $\sigma$-$\pi$ model, as we illustrate
in Fig.~17. In fact, the pattern emerges
 for the same dynamical reason\onlinecite{MandP}:
 the fact that, for $K>1,$ in both the Skyrme model and in the $\sigma$-$\pi$
 model, the phase-shifts associated with ${\sf S}^{L+1}_{11}$ are
much smaller than those of ${\sf S}^{L-1}_{22}$ (cf.~Figs.~15-16).
 We therefore view
it as a model-independent success of the large-$N_c$ approach, whether
one chooses to use skyrmions or explicit baryon fields.

 \subsection{The baryon spectrum of the large-$N_c$ $\sigma$-$\pi$ model}

{}From the partial-wave amplitudes it is easy to extract the baryon resonance
spectrum of the large-$N_c$ $\sigma$-$\pi$ model.
Rather than record when the phase-shifts cross 90 degrees (a crude
criterion sensitive to background potentials), a more robust definition
of a resonance, adopted by experimentalists, is to look for Lorentzian
peaks in the ``speed plots,'' i.e., the plots of
$|dT_{LI_{\rm tot}J_{\rm tot}}/d\omega|$ versus $\omega.$
The  speed plots for a few selected
partial waves are depicted in Fig.~18. Some peaks are unambiguous,
whereas others are admittedly ``in the eye of the beholder,'' but the
same can be said about the corresponding experimental data.

Figure 19 displays the full resonance spectrum obtained in this
fashion, through the $H$-waves ($L=5$), limited to
 what we subjectively consider to be ``two-star" resonances or better.
The step-like structure, in blocks of alternating parity,
is much more pronounced than in
the Skyrme model, and certainly than in Nature. It can be partially
accounted for by noting that, for $L>1,$ the reduced amplitudes
of Fig.~16 dominate those of Figs.~14-15, so that for any
fixed value of $L,$ the resonance location in the four physical
partial-wave amplitudes can be approximated by the resonance location
in the single underlying reduced amplitude ${\sf S}^{L-1}_{22}.$
But this does not explain why the steps
arrange themselves by  definite parity (as we have indicated by the
black bars below the horizontal axis), a feature for which we have no
good understanding.

In general, the resonances in the $\sigma$-$\pi$ model occur at substantially
lower energies than
in the Skyrme model, and in Nature. We have not explored the
parameter space of our model [see Eq.~(\ref{params1})]
in an attempt to rectify this disparity
(as we are confident could be done),
not just because of the computationally-intensive character of these
multi-stage calculations, but also due to the frankly ``toy'' intent
of this model, which we have constructed for illustrative purposes.
We are optimistic that a more realistic model, incorporating the
vector mesons, and properly implementing chiral symmetry, would be in
better agreement with the observed baryon spectrum, while posing no
significant additional conceptual or numerical difficulties beyond those we
have already confronted herein.

The one parameter that we \it have \rm experimented with is the nucleon
size parameter $a_N$, defined in Eq.~(\ref{gaussian}), which acts
as an ultraviolet cutoff. A variation from our nominal value  $a_N=0.52\,$fm
to $a_N=0.60\,$fm shows no discernible effect on the resonance
positions, and only slight changes in the Argand plots themselves,
primarily in the $P$-waves, one of which is shown in Fig.~20.

 \subsection{Some familiar problems}

We have seen that this large-$N_c$ $\sigma$-$\pi$
model (and, we presume, others like it with
explicit baryon fields) shares some notable successes with the Skyrme
model---the Big-Small-Small-Big pattern, the energy-independent
relations between the $I_{\rm tot}=1/2$ and $I_{\rm tot}=3/2$
$\pi N$ amplitudes, the overall richness of the baryon resonance
spectrum, etc. Moreover, the high-energy behavior of the partial
wave amplitudes is much better than in the Skyrme model (see Sec.~V).
Not surprisingly, the $\sigma$-$\pi$ model also shares some of the Skyrme
model's failings. Figure 21 illustrates a specific partial wave
amplitude in the $\sigma$-$\pi$ model, juxtaposed with the experimental
data. Obviously the real-world amplitude is much more inelastic than
the present model. This is because, in the higher partial waves
especially, multiple pion production soon dominates the experimental $\pi N$
amplitudes. Yet, \it formally\rm, processes such as $\pi N\rightarrow
\pi\pi\pi N$ are down by powers of $1/N_c$ compared with $\pi N\rightarrow
\pi N$, and are therefore entirely absent from leading-order theoretical
treatments such as the present paper---as well as from the leading-order
skyrmion treatments\onlinecite{HEHW,Sig,MandK,MandP,Karliner}, which
share the same problem. Below the $\sigma$ threshold, the only source
of inelasticity in the present model is the $\pi\Delta$ channel, exactly
as in the Skyrme model. A theoretical means of summing at least
\it some \rm of the $1/N_c$ corrections, namely those associated with
multiple pion production, would immeasurably improve either approach.

Just as serious is the failure of the $\sigma$-$\pi$ model to bear even
passing resemblance to experiment in the $S$ and $P$ waves. As is
well known, these waves have been the ``Achilles heel'' of the Skyrme
model too. Interestingly, whereas the Skyrme model shows
\it too few \rm resonances in these waves, the $\sigma$-$\pi$ model
errs in the opposite direction: \it too many \rm resonances,
particularly in the $P_{13}$ and $P_{31}$ waves, and including
spurious bound states in the $S_{31}$ and $S_{11}$ channels as
already noted in Sec.~V. The interested reader is referred to
Ref.~\onlinecite{MandP} for a lengthy discussion of the problems
in these lower waves in the Skyrme model,
 which are related, in part, to the failure to
incorporate the translational and (iso)rotational recoil of the
hedgehog (formally $1/N_c$ corrections, but numerically important).
We expect that commentary to apply as well to models
 with explicit baryon fields. For example, the
Weinberg-Tomozawa expression for the
$\pi N$ scattering lengths\onlinecite{WeinTom}, which are predicted by current
algebra, and which dominate the experimental
$S$-wave amplitudes near threshold, formally appear only at
next-leading order in $1/N_c$\onlinecite{MandP}.
This suggests that if one were to start from
an improved effective hadron Lagrangian that respects chiral
symmetry (we remind the reader that the present $\sigma$-$\pi$ model
does \it not\,\rm), and if one were to  calculate to next-leading order
in $1/N_c,$ the most glaring
disagreement with experiment in the $S$-waves ought to be repaired.
Fixing the $P$-waves will require, at the least, ($i$) the splitting
of the $\Delta$ from the nucleon (again, a $1/N_c$ effect), and
($ii$) the incorporation of the Compton-type diagrams,
particularly Figs.~1a and 1b, the ameliorating effect of which has already
been examined in the Skyrme model\onlinecite{DiakPet,Japs}.

\acknowledgments

We acknowledge valuable input from many of our colleagues, most notably
Peter Arnold, Charles Benesh, Nick Dorey,  Jim Friar, Terry Goldman,
Gerry Hale,  Jim Hughes, Marek Karliner,  Arthur Kerman,  Wim Kloet,
Jim McNeil, Charles Price, Rob Timmermanns, and John Tjon. We also
thank Aneesh Manohar for commenting on the draft.

This work has been supported by
 the Division of High Energy and Nuclear Physics,
 Energy Research, Department of Energy.  MPM has also benefitted from
an SSC Fellowship during part of the time we have been
working on this problem.

% *** References ***

\pagebreak
\appendix
\section{Numerical and Computational Details}

\subsection{Solving the coupled non-linear equations}

We describe here  the method we  use for solving the non-linear
 classical field equations, Eq.~(\ref{FGeqns}),
 for the full $\sigma$-$\pi$ model.
(Exactly the same technique is used for the $\sigma$-only and $\pi$-only
warm-up problems discussed before that point.)
These are coupled ordinary nonlinear
differential equations. The nonlinearity implies that
 there is some sensitivity in finding solutions; indeed,
for some ranges of parameters, one may not be able to find solutions at all.

We are looking for solutions of these equations
which are regular at the origin
and which fall off asymptotically at large distances.
Examining the indicial behavior of Eq.~(\ref{FGeqns})
near $r$ = 0, we find that
\begin{eqnarray}
  F(0) & = & 0 \ , \quad \quad  F'(0) = A \ , \quad \quad
  G(0)  =  B  \quad \quad  G'(0) = 0 \ .
\end{eqnarray}
The indicial values $A$ and $B$, along with asymptotic scale parameters $C$ and
$D$ defined below in Eqs.~(\ref{asympF}) and (\ref{asympG}), are initially
unknown.
Their values will be fixed by the non-linearity when we solve the differential
equations.

As $r \to \infty$, the pion profile function is required to have the usual
Yukawa-like fall-off,
\begin{equation}
   F(r) \to C \exp(-m_\pi r) /r \ .
   \label{asympF}
\end{equation}
Things are more complicated for $G(r)$ as $r \to \infty$ because the asymptotic
behavior of $G(r)$ is governed by the coupling term to two pions.
That is, it will fall off like $\exp(-2m_\pi r)/r^2$ rather than the faster
$\exp(-m_\sigma r)/r$.
After some analysis (solving a linearized Eq.~(\ref{Geqn}) using a Green's
function technique with the $F^2(r)$ term on the right-hand-side providing
the inhomogeneity),
\begin{equation}
   G(r) \to D {e^{-m_\sigma r} \over r}
        \ - \  {\lambda v C^2 \over 2 m_\sigma}\;
          {e^{-2m_\pi r}  \over r} E_1[-(m_\sigma + 2m_\pi) r] \ ,
   \label{asympG}
\end{equation}
where $E_1$ is an exponential integral \cite{AandS}.

The solution of Eq.~(\ref{FGeqns}) constitutes a two-boundary-value problem.
We solve the equations using a standard shoot and match technique
\cite{NumRec}.
Not knowing, at first, the values for the  ``scale parameters'' $A \ldots D$,
we make an initial guess for their values and proceed to refine them with an
iterative procedure.

The procedure is as follows.
Given $A \ldots D$, we integrate out from the origin using a Runge-Kutta
technique to our matching radius, $r_m$, which we choose to be $a_N$.
We then Runge-Kutta integrate backwards to $r_m$ from $r_a$, a point where the
asymptotic forms shown above are expected to hold.
We typically choose $r_a$  to be 2 fm.
The values at $r_m$ of $F$ and $G$, and their derivatives, from the two
integrations are then compared.
The mismatches, or discontinuities, give four conditions which can be used to
choose corrections to the guessed values of $A \ldots D$ that would drive the
discontinuities toward zero.
(This is a generalization to four variables of the Newton-Rapheson method;
it requires four more passes of Runge-Kutta integrations to compute the
derivatives of the discontinuities with respect to the scale parameters.)
Correcting the $A \ldots D$ as calculated, one can repeat this process,
hopefully getting a better, less discontinuous $F(r)$ and $G(r)$.
The procedure 	is iterated until it converges to a solution.

The above iterative procedure was programmed in Fortran and was originally run
as a batch job on a VAX-VMS minicomputer from a terminal command line.
For small values of $\lambda$ (i.e., small non-linear contributions) the
program would converge reasonably well.
However, as the non-linearity was increased, corresponding to values of
$\lambda$ given by Eq.~(\ref{params}), the convergence became
more delicate.
Thus, we found it useful to make the code more interactive, so that the user
could watch plots of $F$ and $G$ at every stage of the iteration and, if the
process were going astray, stop it and start again with a new set of starting
scale parameters.
The computing was transferred to a NeXT workstation and a NeXTSTEP front end to
the Fortran program was developed \cite{frontfort}.
With this front end to aid the user, it was much easier finding a solution for
$\lambda$ = 91.

\subsection{Runge-Kutta Approach to the Scattering Differential Equations}

The quantum scattering phase shifts in, say, the $\sigma$-only model of
Sec.~II, is given by the asymptotic form of the radial wave function determined
by Eq.~(\ref{sigscat}).
This linear differential equation (and the coupled-channels variants of
Secs.~III and V) are readily solved by Runge-Kutta integration \cite{NumRec},
simply by integrating out from the origin.
Starting values are taken from the regular solution for
the partial wave of the angular momentum $K$ under consideration,
i.e., with behavior like  ${\hat{\jmath}}_K(qr) \propto r^{K+1}$.
(The scale can be chosen arbitrarily because the equation is linear.)

At a large enough distance, typically 10 fm or so,
the solution is well-approximated by a
linear combination of $\hat{\jmath}_K(qr)$ and $\hat n_K(qr)$.
Fitting the coefficients of these
 functions, one can then form the $S$-matrix element and
thence compute the phase shift.
For the coupled-channels case, the $S$-matrix is found using the
procedure given, for example, in Ref.~\onlinecite{MandK}  utilizing matrices
formed from these fitted coefficients.

\subsection{Comments on Solving the Integral Scattering Equations}

Numerically, Eq.~(\ref{IntEqn})
is solved using the Nystrom method \cite{NumRec}.
For our three-rowed column vector, this
procedure involves the inversion of a large $3 N \times 3 N$ matrix, where
$N$ is the number of meshpoints.
The time to do that inversion goes like $N^3$.
We found that reasonable accuracy (1\% or better) in the extracted $S$-matrix
elements is obtained with $N = 150$ mesh points.
These are typically distributed as follows: 120 mesh points, evenly spaced,
from $r$ = 0 to 3 fm and 30 points, evenly spaced, from 3 to 10 fm.

In debugging the coding we found it very useful to verify that the $S$-matrix
is not only unitary but also symmetric (a consequence of time reversal
invariance).
Another useful check on the code was to see that the scattering wave functions
and $S$ are continuous through the $\sigma$ threshold and agree with the
prediction of the Runge-Kutta method above the threshold.

\section{Figure Captions}

\begin{enumerate}

%piNcompton.eps
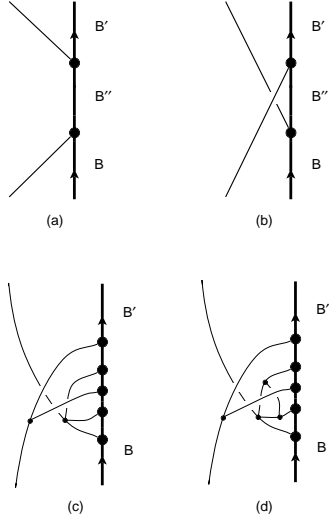
\item{Compton-type graphs contributing to meson-baryon
scattering. Henceforth, directed lines stand for baryons, all
other lines are mesons. The baryons $B$, $B'$ and $B''$ stand for three members
of the $I=J$ tower of baryons. Fig.~1(d) contains a purely mesonic loop and
so is subleading in $1/N_c$; Figs.~1(a)-(c) do not,
and so contribute at leading order [9].
}

%piNmesonExch.eps
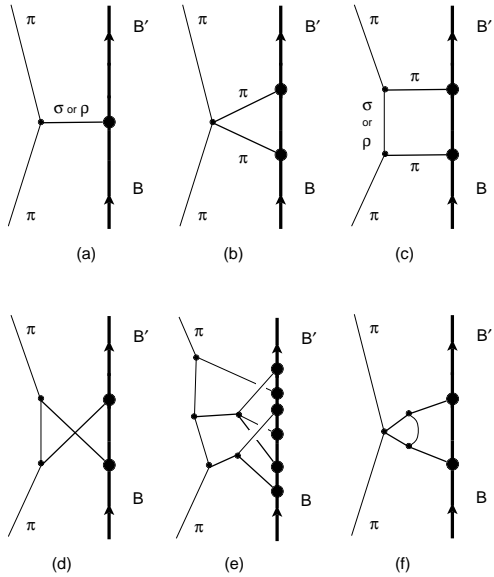
\item{Exchange-type graphs contributing to meson-baryon
scattering.  Fig.~2(f) contains a purely mesonic loop and so is subleading
in $1/N_c$; Figs.~2(a)-(e) do not, and so contribute at leading order.
In this paper we explicitly sum all such leading-order graphs in
a model containing both $\pi$ and $\sigma$ mesons.
}

%NNmesonExch.eps
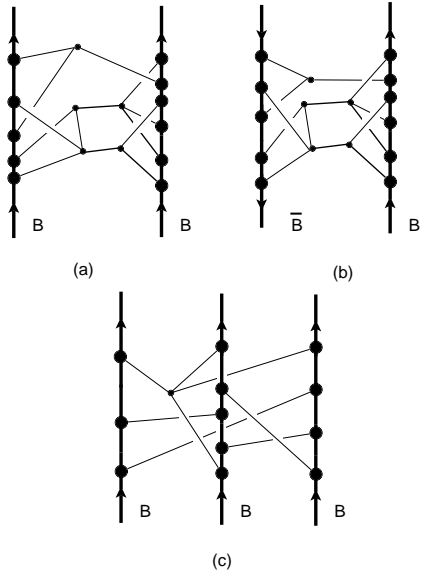
\item{Examples of exchange-type graphs contributing to
(a) the baryon-baryon interaction, (b) the baryon-antibaryon interaction,
and (c) the baryon-baryon-baryon interaction. As none of these
particular graphs contains a purely mesonic loop, they all contribute
to leading order in $1/N_c,$ and can be summed using the methods
of Ref.~[9].
}

%nMesonExch.eps
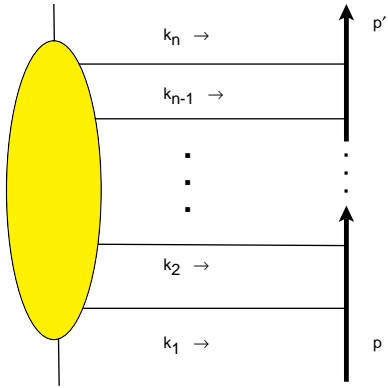
\item{An uncrossed $n$-meson exchange graph included in the
summation of Fig.~2. The shaded blobs contains meson
self-interactions that do not concern us.
}

%4MesonsCrossed.eps
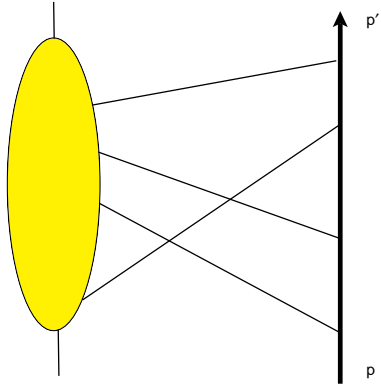
\item{A specific ``crossing'' or ``tangling'' of Fig.~4, for $n=4$.
There are $n!$ such tanglings.
}

%nMesonSources.eps
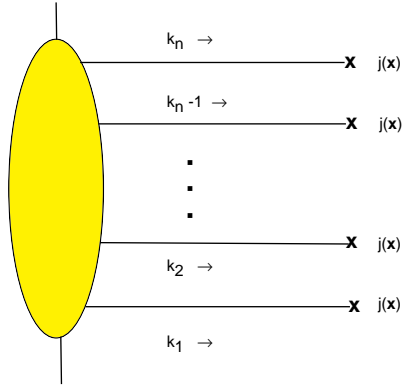
\item{The sum of all $n!$ tanglings including Figs.~4-5.
In all three figures, the contents of the shaded blob are held constant.
classical sources.
}

%sigClassExpsn.eps
\item{The tree-level one-point function $\sigma_{\rm cl},$
where the baryon has been replaced by an external source $j({\bf x})$.
}

%sigQuantExpsn.eps
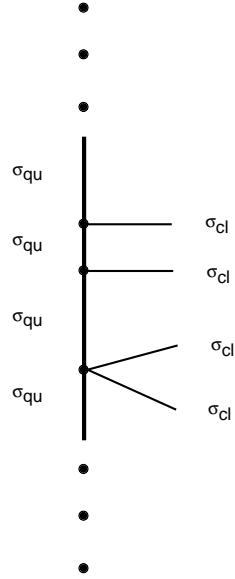
\item{The fluctuating field $\sigma_{\rm qu}$ propagating
through the nontrivial background generated by $\sigma_{\rm cl}.$
Vertices can be read off from the quadraticized Lagrangian, Eq.~(15).
}

%sigOnlyG.eps
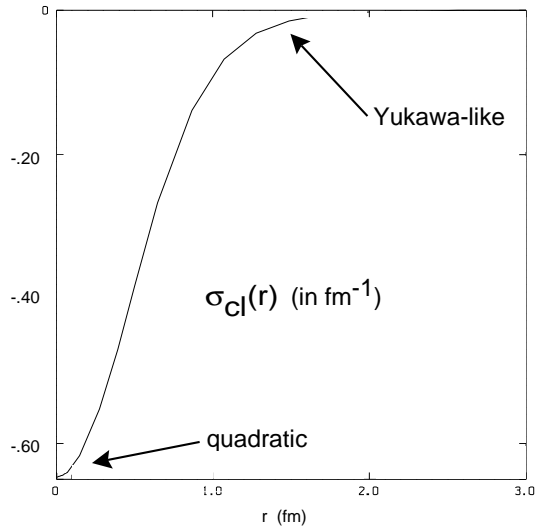
\item{The profile function $\sigma_{\rm cl}(r)=G(r)$
for the $\sigma$-only model of Sec.~II.
}

%sigphases.eps
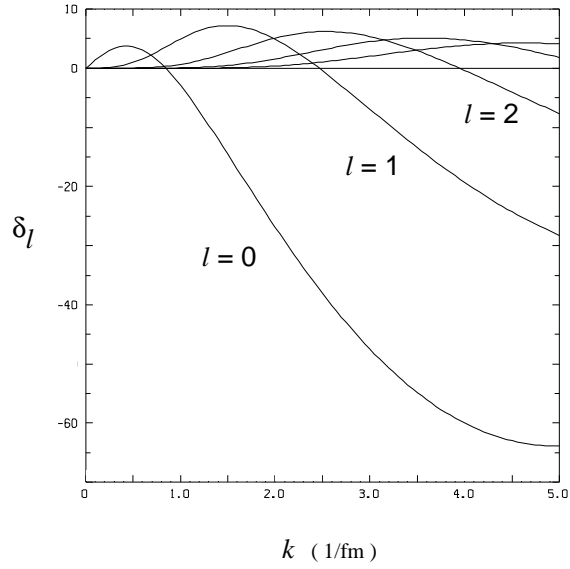
\item{Phase shifts for the $\sigma$-only model, as a function
of $\sigma$ momentum, for model parameters $m_\sigma=600\,$MeV,
$g=13.6$, and $a_N=0.5\,$fm.
}

%piOnlyF.eps
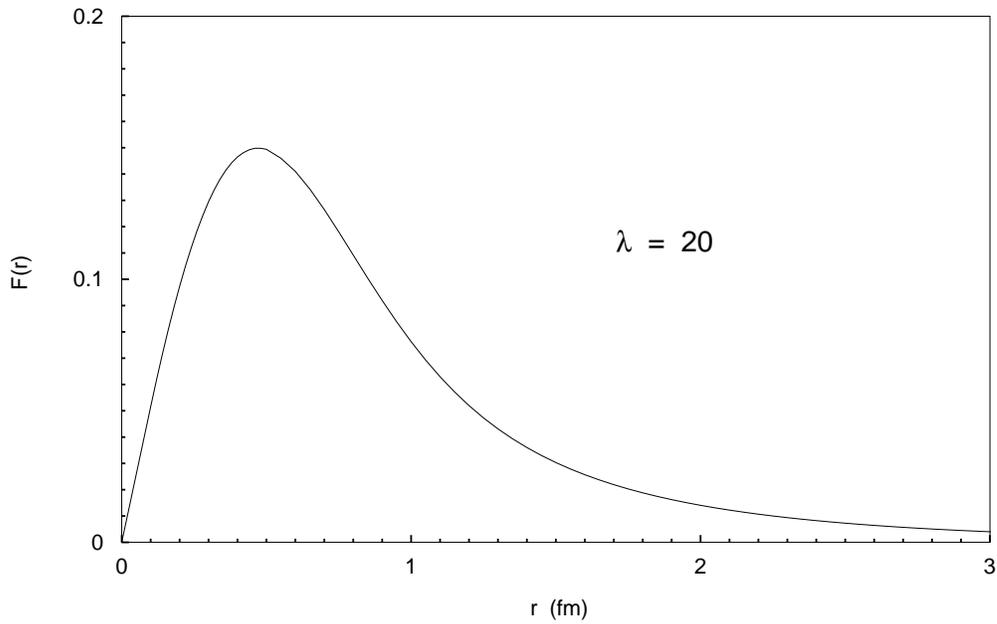
\item{Classical profile $F(r)$ in inverse fermis
for the $\pi$-only model of Sec.~III.
}

%FandG.eps
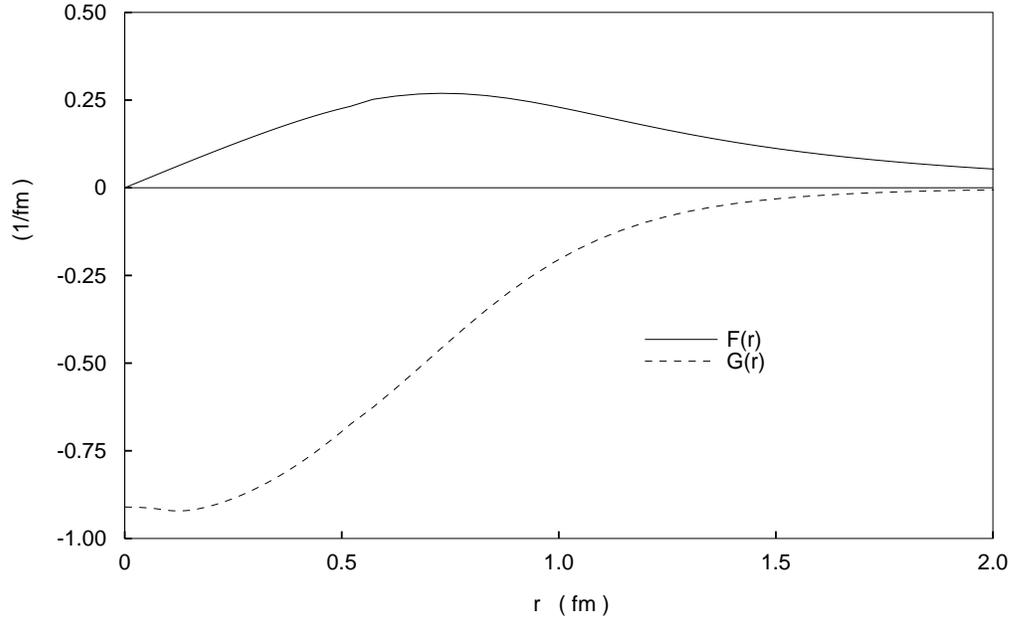
\item{Classical profile functions $F(r)$ and $G(r)$
for the $\sigma$-$\pi$ model, with parameters as in text.
}

%potnls.eps
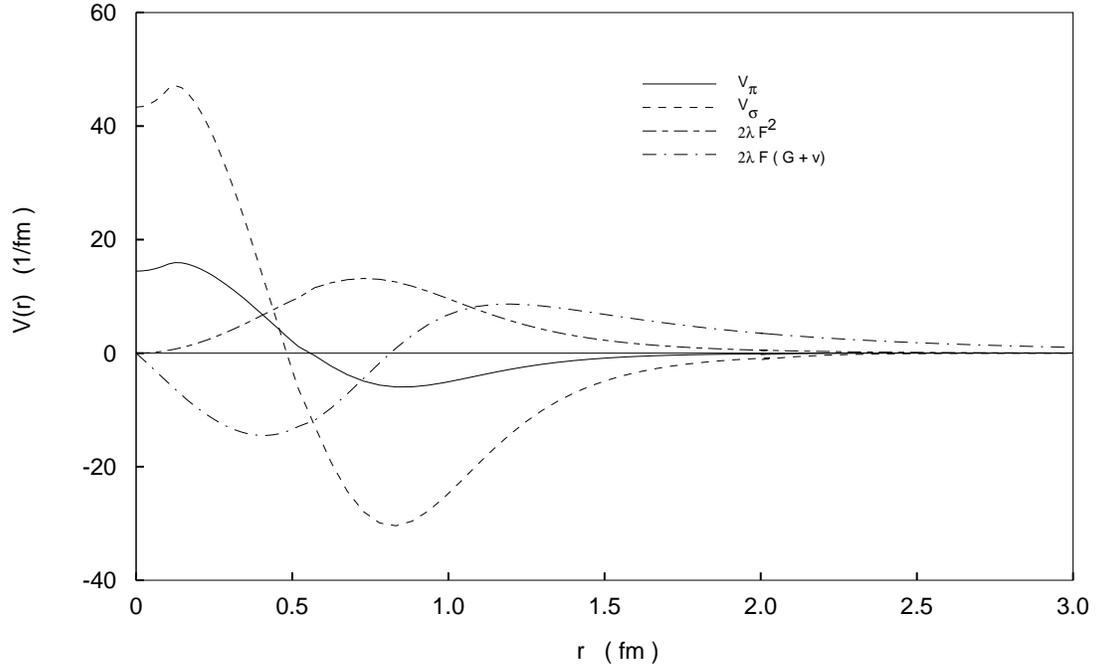
\item{Diagonal and off-diagonal potentials appearing
in the small-fluctuations equations.
}

%delK0
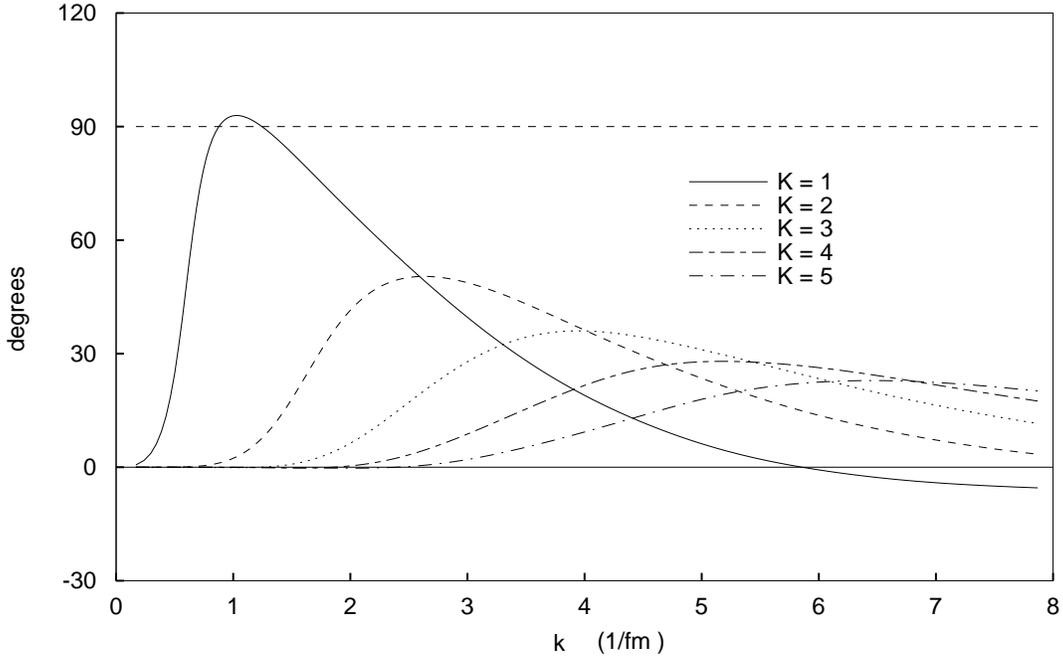
\item{The phase shifts $\delta_K$ plotted against
meson momentum $k$ in the baryon rest frame (recall that baryon
recoil is subleading in $1/N_c$).
}

%delKmi
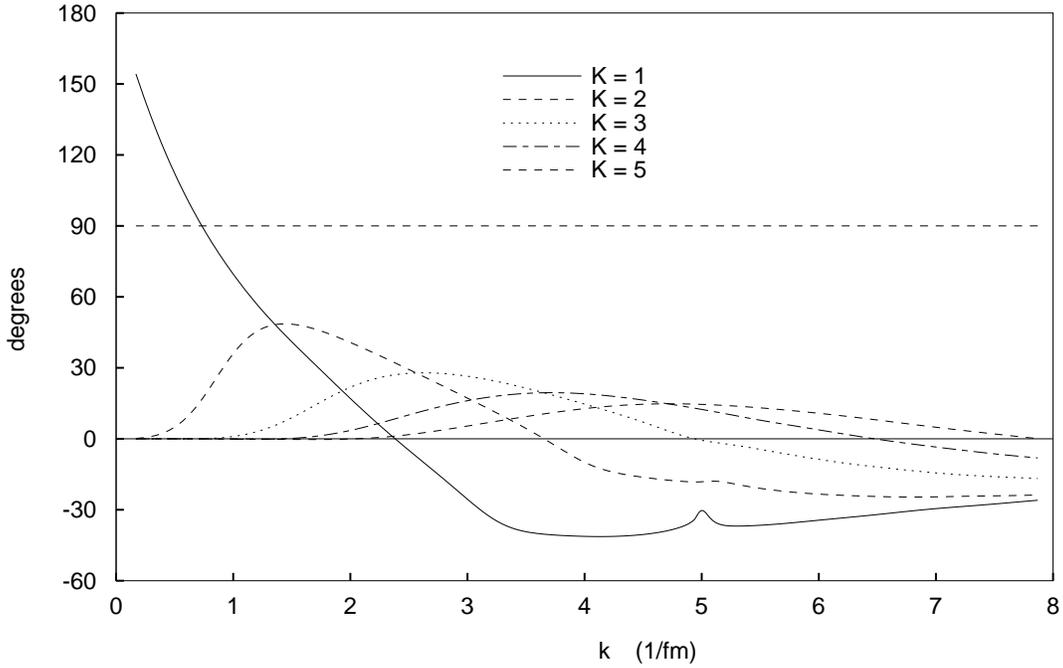
\item{The phase shifts $\delta^K_{11}$ plotted against
meson momentum $k$ in the baryon rest frame.
}

%delKpl
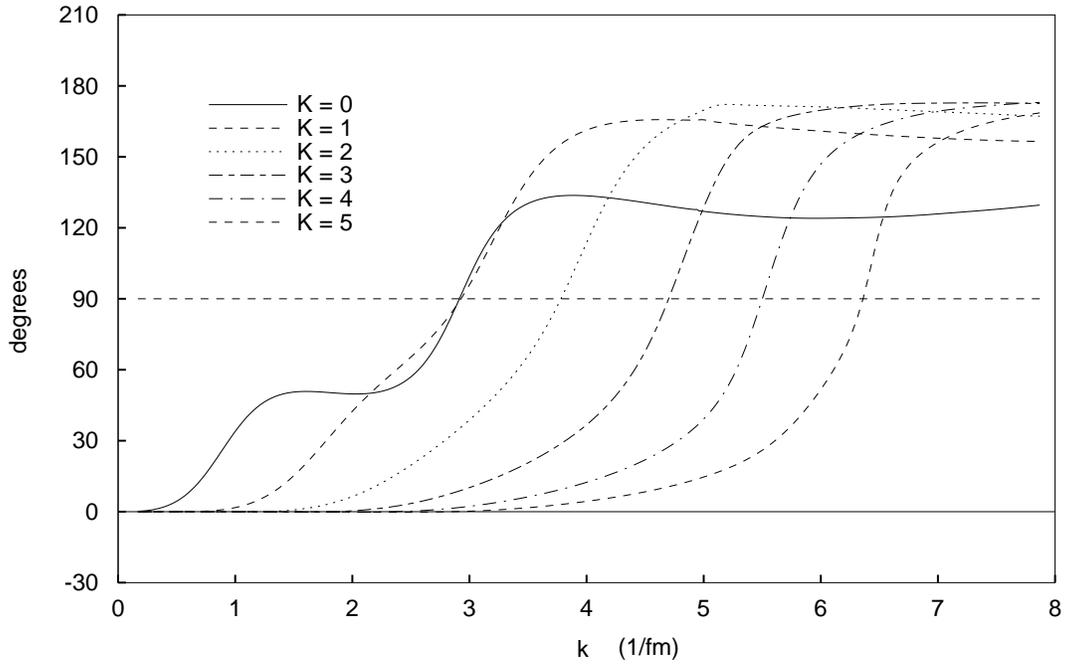
\item{The phase shifts $\delta^K_{22}$ plotted against
meson momentum $k$ in the baryon rest frame.
}

%Fwaves.eps
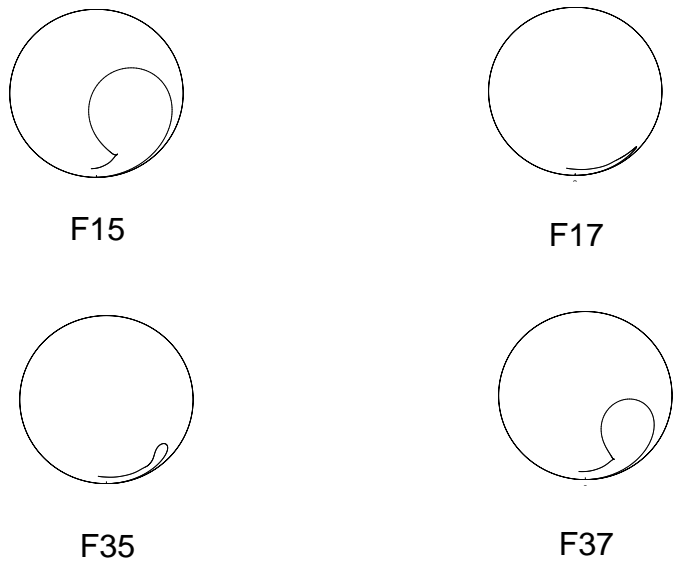
\item{Argand plots for
the four $F$-wave elastic $\pi N$ partial wave amplitudes, illustrating
the ``Big-Small-Small-Big'' pattern (the $F_{15}$ and $F_{37}$ are
big, the others small).
}

%speedT3Lmi.eps
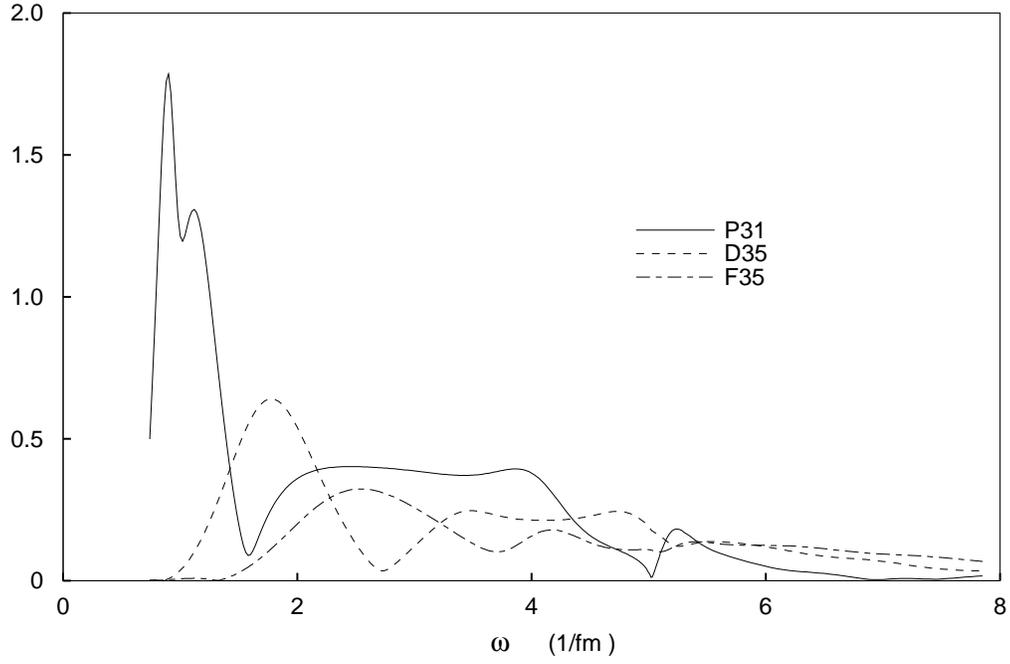
\item{Examples of speed plots for selected $\pi N$
amplitudes, from which resonance positions are extracted.
}

%bands.eps
\item{
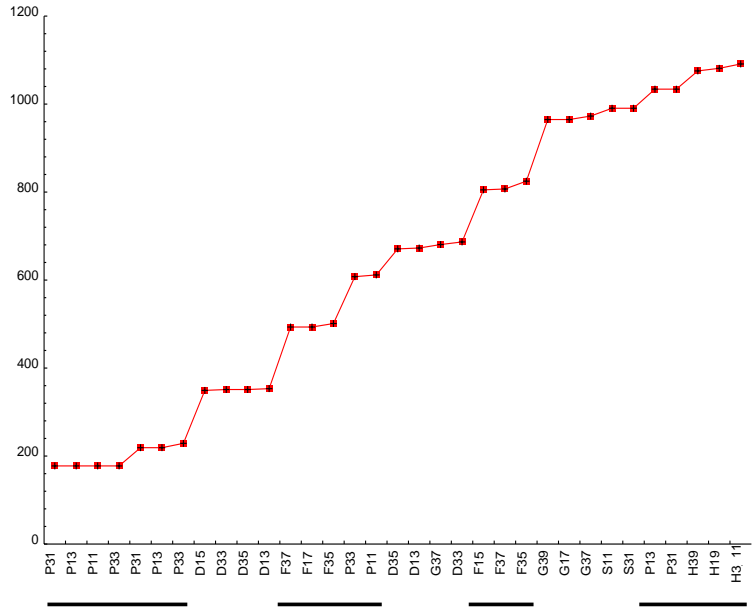
The baryon spectrum of the $\sigma$-$\pi$ model.
The vertical axis measures excitation energy in MeV above the
nucleon/hedgehog mass. Not pictured are the two spurious bound states
in the $S_{11}$ and $S_{31}$ channels.
}

%dblargP13.eps
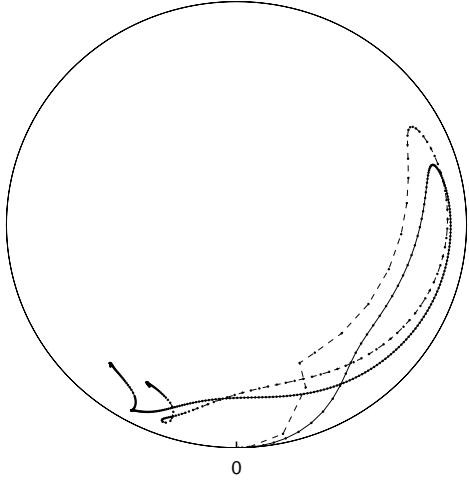
\item{Dependence of the $P_{11}$ amplitude on the nucleon size parameter $a_N,$
for  $0.52$ and $0.60$ fm (solid and dashed lines, respectively).
}

%G17data.eps
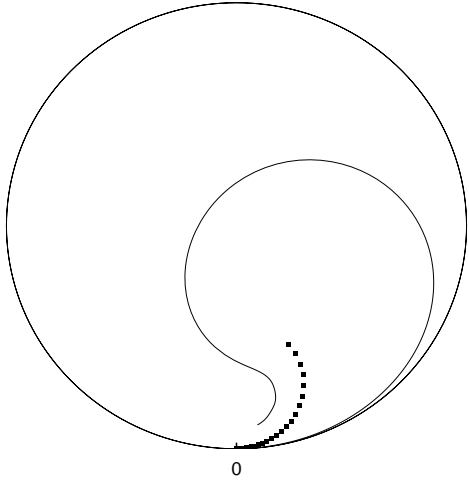
\item{Comparison of $G_{17}$ amplitude between the
$\sigma$-$\pi$ model (solid line) and experiment (dotted line).
The pion kinetic energy for each curve ranges from 0 to 1600 MeV.
}

\end{enumerate}

\end{document}